\documentclass[a4paper,11pt]{article}

\usepackage[percent]{overpic}
\usepackage{jcappub} 

\usepackage{graphics}
\usepackage{longtable}

\usepackage{amssymb}

\usepackage{amsmath}

\title{\boldmath Search of Dark Matter Annihilation in the Galactic Centre using the ANTARES Neutrino Telescope}

\author[a]{S.~Adri\'an-Mart\'inez}
\author[b]{A.~Albert}
\author[c]{M.~Andr\'e}
\author[d]{G.~Anton}
\author[a]{M.~Ardid}
\author[e]{J.-J.~Aubert}
\author[f]{B.~Baret}
\author[g]{J.~Barrios-Mart\'{\i}}
\author[h]{S.~Basa}
\author[e]{V.~Bertin}
% ...

\affiliation[a]{\scriptsize{Institut d'Investigaci\'o per a la Gesti\'o Integrada de les Zones Costaneres (IGIC) - Universitat Polit\`ecnica de Val\`encia. C/  Paranimf 1 , 46730 Gandia, Spain.}}
\affiliation[b]{\scriptsize{GRPHE - Universit\'e de Haute Alsace - Institut universitaire de technologie de Colmar, 34 rue du Grillenbreit BP 50568 - 68008 Colmar, France}}
\affiliation[c]{\scriptsize{Technical University of Catalonia, Laboratory of Applied Bioacoustics, Rambla Exposici\'o,08800 Vilanova i la Geltr\'u,Barcelona, Spain}}
\affiliation[d]{\scriptsize{Friedrich-Alexander-Universit\"at Erlangen-N\"urnberg, Erlangen Centre for Astroparticle Physics, Erwin-Rommel-Str. 1, 91058 Erlangen, Germany}}
\affiliation[e]{\scriptsize{CPPM, Aix-Marseille Universit\'e, CNRS/IN2P3, Marseille, France}}
\affiliation[f]{\scriptsize{APC, Universit\'e Paris Diderot, CNRS/IN2P3, CEA/IRFU, Observatoire de Paris, Sorbonne Paris Cit\'e, 75205 Paris, France}}
\affiliation[g]{\scriptsize{IFIC - Instituto de F\'isica Corpuscular, Edificios Investigaci\'on de Paterna, CSIC - Universitat de Val\`encia, Apdo. de Correos 22085, 46071 Valencia, Spain}}
\affiliation[h]{\scriptsize{LAM - Laboratoire d'Astrophysique de Marseille, P\^ole de l'\'Etoile Site de Ch\^ateau-Gombert, rue Fr\'ed\'eric Joliot-Curie 38,  13388 Marseille Cedex 13, France}}
% ...

\abstract{A search for high-energy neutrinos coming from the     
direction of the Galactic Centre is performed using the data recorded by 
the ANTARES neutrino telescope from 2007 to 2012. The event selection 
criteria are chosen to maximise the sensitivity to possible signals produced by   
the self-annihilation of weakly interacting massive particles accumulated around the centre of 
the Milky Way with respect to the atmospheric background. After data unblinding, the number of 
neutrinos observed in the line of sight of the Galactic Centre is found to be compatible with 
background expectations. The 90\% C.L. upper limits in terms of the neutrino+anti-neutrino flux, 
$\rm \Phi_{\nu_{\mu}+\bar{\nu}_\mu}$, and the velocity averaged annihilation cross-section, $\rm <\sigma_{A}v>$, 
are derived for the WIMP self-annihilation channels into $\rm b\bar{b},W^{+}W^{-},\tau^{+}\tau^{-},\mu^{+}\mu^{-},\nu\bar{\nu}$. 
The ANTARES limits for $\rm <\sigma_{A}v>$ are shown to be the most stringent for a neutrino telescope over the WIMP masses 
$\rm 25\,GeV < M_{WIMP} < 10\,TeV$.}


\begin{document}
\maketitle
\flushbottom

\section{Introduction}
\label{introduction}

Observations in cosmology and astrophysics indicate that about $84$\% of the matter 
in the Universe, the so-called dark matter, is non-baryonic, non-relativistic, and does interact 
only through gravity.~\cite{darkmatter,dmschumann,dmsalati,pdg2012}. 

These observations involve the internal dynamics of galaxy clusters~\cite{galclust}, 
the rotation curves of galaxies~\cite{galrot}, weak lensing~\cite{bulclust}, also the Cosmic Microwave 
Background (CMB) from which the relic density of cold dark matter (CDM) in the Universe is at present deduced to be 
$\rm \Omega_{CDM}h^{2} = 0.1199 \pm 0.0022$ (Planck+WMAP)~\cite{planck2015,wmap7yr}. 

A common assumption is that dark matter is made of Weakly Interacting Massive 
Particles (WIMPs) that form halos in which the visible baryonic part of galaxies is embedded. There are 
a variety of candidates for WIMPs, among which those provided by theories based on 
supersymmetry (SUSY) attract a great deal of interest. In some classes of minimal supersymmetric (MSSM) and minimal 
universal extra-dimensional (mUED) extensions of the Standard Model (SM), the lightest particle (LP) is stable thanks to 
the conservation of a model-dependent parity that forbids its decay into SM particles. Consequently, 
these LPs can only annihilate in pairs, making them a possible WIMP candidate for dark matter~\cite{lsp,lkp}. 
In these models, secondary high-energy neutrinos are produced from the decay of the LPs' self-annihilation products. 
In mUED models, neutrinos can even be produced directly in the self-annihilations, since there is no helicity suppression 
of fermion pair production. 

The search for WIMPs can be performed either directly by recording the recoil energy of 
nuclei when WIMPs scatter off them in suitable detectors, or indirectly. The indirect approach, 
which is adopted here, exploits a radiation signature (gamma-ray, synchroton, positron, anti-proton 
or, as in this case, neutrino flux) produced by the self-annihilation of WIMPs accumulated in massive astrophysical objects 
such as the Galactic Centre (GC), the Sun or the Earth~\cite{indirectdm}.

For the case of the GC, dealt with in this paper, where the density of dark matter in the galactic halo is supposed to be 
the highest, WIMPs self-annihilate to SM particles whose decay or hadronisation (if not directly to neutrinos) give rise to the 
production of high energy neutrinos which can travel from the GC to the Earth and be detected by neutrino telescopes. 

\bigskip
In this paper, the indirect search for dark matter by looking for high-energy neutrinos coming from the GC, 
using the 2007-2012 data recorded by the ANTARES neutrino telescope, is described. The layout of the paper 
is as follows. In Section~\ref{antares}, the main features of the ANTARES neutrino telescope 
and the reconstruction algorithms used in this work are explained. In Section~\ref{simulation}, 
the expected signal from WIMP self-annihilation from the GC, and the background expected from atmospheric 
muons and neutrinos are reported. In Section~\ref{optimisation}, the method used to optimise the selection of 
the neutrino events is described. Finally, the results obtained are discussed in Section~\ref{results}, where limits 
on the neutrino plus anti-neutrino flux $\rm \Phi_{\nu_{\mu}+\bar{\nu}_\mu}$ are derived from the absence of a signal. 
The corresponding 90\% C.L. upper limits on the velocity-averaged annihilation cross-section $\rm <\sigma_{A}v>$ are 
obtained for different benchmark channels of self-annihilation, and compared to the latest constraints from other 
experiments. In the following, neutrino will mean neutrino plus anti-neutrino, unless explicitly stated otherwise.   


\section{The ANTARES Neutrino Telescope}
\label{antares}

ANTARES is the first undersea neutrino telescope and the largest of its kind in the Northern 
Hemisphere~\cite{antares}. It is located at 2475 m below the Mediterranean Sea level, $40$ km 
offshore from Toulon (France) at $42^{\circ} 48$' N and $6^{\circ} 10$' E. The telescope 
consists of $12$ detection lines with 25 storeys each. A standard storey includes 
three optical modules (OMs)~\cite{OM}, each housing a 10-inch photomultiplier~\cite{PMT} and a local control 
module that contains the electronics~\cite{frontend, DAQ}. The OMs are orientated 45$^{\circ}$ 
downwards in order to optimise their acceptance to upgoing light and to avoid the effect of 
sedimentation and biofouling~\cite{biofouling}. The length of a line is 450 m and the horizontal 
distance between neighbouring lines is 60-75 m. In one of the lines, the upper storeys are dedicated 
to a test system for acoustic neutrino detection~\cite{amadeus}. Similar acoustic devices are also 
installed in an additional line that contains instrumentation aimed to measure environmental 
parameters~\cite{instrumentation}. The location of the active components of the lines is known 
to better than 10~cm by a combination of tiltmeters and compasses in each storey and a series of
acoustic transceivers (emitters and receivers) in certain storeys along the line and surrounding 
the telescope~\cite{alignment}. A common time reference is maintained in the full detector by means 
of a 25 MHz clock signal broadcast from shore. The time offsets of the individual optical modules 
are determined in dedicated calibration facilities onshore and regularly monitored in situ by means 
of optical beacons distributed at various points of the apparatus which emit short light pulses 
through the water~\cite{OBs}. This allows a sub-nanosecond accuracy on the relative 
timing~\cite{timing}. Additional information on the detector can be found in Ref.~\cite{antares}.

Data-taking started with the first 5 lines of the detector installed in 2007. The full detector 
was completed in May 2008 and has been operating continuously ever since, except for some short periods 
in which repair and maintenance operations have taken place. Other physics results using this data-taking 
period can be found elsewhere~\cite{diffuse,monopole,nuosc,ps,grb,atmnu,dmsun,fermibub,pe}.

High-energy muon neutrinos interacting in the matter before the detector produce relativistic 
muons that can travel hundreds of metres and cross the detector or pass nearby. These muons induce 
Cherenkov light when travelling through the water, which is detected by the OMs. From the time and 
position information of the photons provided by the OMs, the direction of the muons, which is well 
correlated with that of the neutrinos, is reconstructed.

\begin{figure}[!t]
\begin{center}
\includegraphics[width=0.8\linewidth]{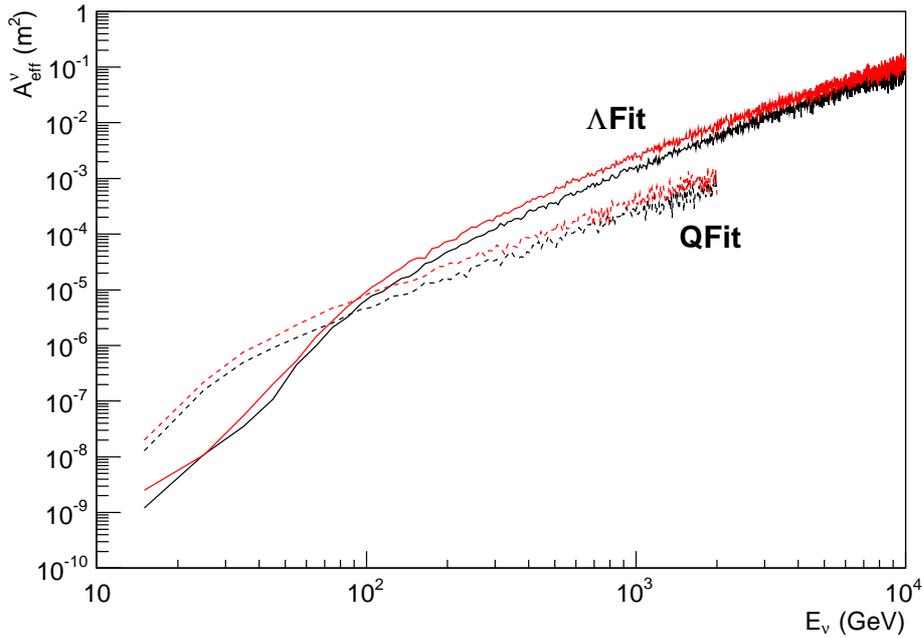}
\caption{Effective area to muon neutrinos (black) and anti-neutrinos (red), $\rm A_{eff}^{\nu}$ (m$\rm ^2$), for $\rm QFit$ (dashed line) and $\rm \Lambda Fit$ (solid line), 
for the 12 line configuration of the detector. The $\rm QFit$ is limited here to its energy range of relevance, $\rm E_{\nu} \leq 2$ TeV.}
\label{aefffig}
\end{center}
\end{figure}

\bigskip
Two reconstruction algorithms are used in this paper. The first one is based on the minimisation of 
a $\rm \chi^{2}$-like quality parameter of the reconstruction, $\rm Q$, which uses the difference between the expected 
and measured times of the detected photons, taking into account the effect of light absorption in the water~\cite{bbfit}. The second 
algorithm consists of a multi-step procedure to fit the direction of the muon track by maximising a likelihood ratio, 
$\rm \Lambda$, which describes the quality of the reconstruction~\cite{ps}. In addition to the $\rm \Lambda$ 
parameter, the uncertainty of the muon track angle, $\rm \beta$, is used for the track selection. 
These two algorithms are respectively called here $\rm QFit$ and $\rm \Lambda Fit$. $\rm QFit$ is used for muon events 
reconstructed in a single detection line (single-line events), and $\rm \Lambda Fit$ for muon events reconstructed on more 
than one detection line (multi-line events) in order to reach the best efficiency of reconstruction in the entire neutrino energy range. 
For $\rm QFit$ and $\rm \Lambda Fit$, a selection on the quality parameter 
$\rm Q < 0.8$ and the couple ($\rm \Lambda > -5.7$; $\rm \beta < 0.5^{\circ}$) have been used, respectively. These values of quality parameters 
are extracted from the optimisation step detailed in Section~\ref{optimisation}. $\rm QFit$ and $\rm \Lambda Fit$ present different efficiencies 
of reconstruction, characterised by the effective areas for muon neutrinos, $\rm A_{eff}^{\nu}$ ($\rm m^2$), which are shown in Figure~\ref{aefffig} as 
a function of the energy of the primary neutrinos $\rm E_{\nu}$ (GeV). An effective area, $\rm A_{eff}^{\nu}$, at a given energy is defined as the ratio 
between the neutrino event rate ($\rm s^{-1}$) in a detector and the neutrino flux ($\rm m^{-2} \cdot s^{-1}$) at that energy. As one can see, the $\rm QFit$ 
reconstruction strategy has a larger efficiency in the low energy regime with $\rm A_{eff}^{\nu}$ higher than the one obtained with $\rm \Lambda Fit$ for 
$\rm E_{\nu} < 100$ GeV.

\begin{figure}[!t]
\begin{center}
\includegraphics[width=0.8\linewidth]{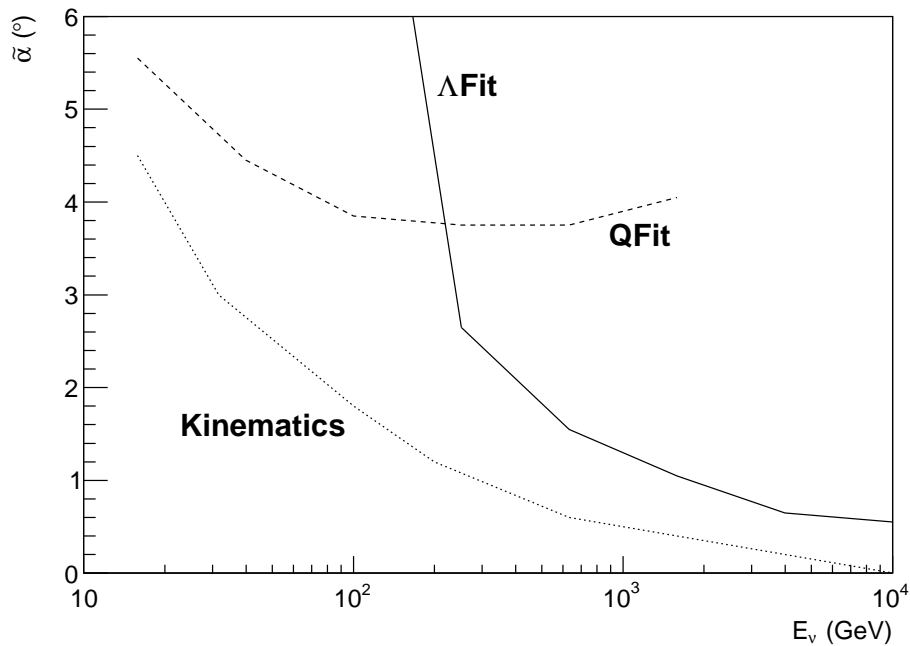}
\caption{Median angular resolution, $\rm \tilde{\alpha}$ ($^{\circ}$), on the muon neutrino direction, for the complete 12 line 
configuration of the detector. The performances for $\rm QFit$ (dashed line) and $\rm \Lambda Fit$ (solid line) are illustrated, 
as is the kinematic counterpart (dotted line), the median angle between the primary neutrino and the outgoing muon at the vertex 
of interaction. For $\rm QFit$, $\rm \tilde{\alpha}$ is defined as the median resolution on the zenith angles $\rm \theta_{rec,nu}$ 
(see text for details).}
\label{angresfig}
\end{center}
\end{figure}

Beside the effective area, $\rm QFit$ and $\rm \Lambda Fit$ yield different median angular resolutions, 
$\rm \tilde{\alpha} = med[\mid\arccos(\vec{d}_{rec}\cdot\vec{d}_{nu})\mid]$ ($^{\circ}$) between the reconstructed muons, 
$\rm \vec{d}_{rec}$, and the corresponding primary neutrinos, $\rm \vec{d}_{nu}$, as shown in Figure~\ref{angresfig}. For $\rm QFit$, 
since in single-line events only the zenith angle, $\rm \theta$, is reconstructed, $\rm \tilde{\alpha}$ is defined as the median 
of the difference between $\rm \theta_{rec}$ and $\rm \theta_{nu}$, $\rm \tilde{\alpha} = med[\mid\theta_{rec}-\theta_{nu}\mid]$. 
$\rm \Lambda Fit$ yields a median angular resolution $\rm 6^{\circ} > \tilde{\alpha} > 0.5^{\circ}$ for the primary neutrino energy range 
$\rm 100\,GeV < E_{\nu} < 10\,TeV$, whilst $\rm QFit$ reaches a $\rm 5.5^{\circ} > \tilde{\alpha} > 3.8^{\circ}$ for $\rm 15\,GeV < E_{\nu} < 1\,TeV$. 


Both reconstruction algorithms are used in the search and the one with the best performance in a given energy range is selected to obtain 
the best possible limits in that range. 

\section{Signal and background simulation}
\label{simulation}

The energy spectrum of muon neutrinos arriving at the Earth's surface from WIMP self-annihilation occurring in the GC's vicinity is computed using 
Ref.~\cite{cirelli}. The muon neutrinos resulting from the self-annihilation channels in SM particles are propagated to the Earth for 17 different WIMP 
masses in the range from 25~GeV to 10~TeV. To compute the initial energy spectra at the GC, five benchmark self-annihilation channels of WIMP-like 
dark matter particles into SM particles are used:

\begin{equation}
\rm{WIMP\,WIMP \rightarrow b\bar{b},W^{+}W^{-},\tau^{+}\tau^{-},\mu^{+}\mu^{-},\nu_{\alpha}\bar{\nu}_{\alpha}} \, .
\label{benchch}
\end{equation}

Also, neutrinos flavours, $\rm \alpha = e,\mu,\tau$, can be produced directly or as subsequent decay products of the SM particles listed above. 
The propagation of the neutrinos includes the three-flavour neutrino oscillations in vacuum along the line of sight from the GC to the 
surface of the Earth using the values from Table 8 in Ref.~\cite{oscs}. The muon neutrino energy spectrum can be written as: 

\begin{figure}[!t]
\begin{center}
\includegraphics[width=0.8\linewidth]{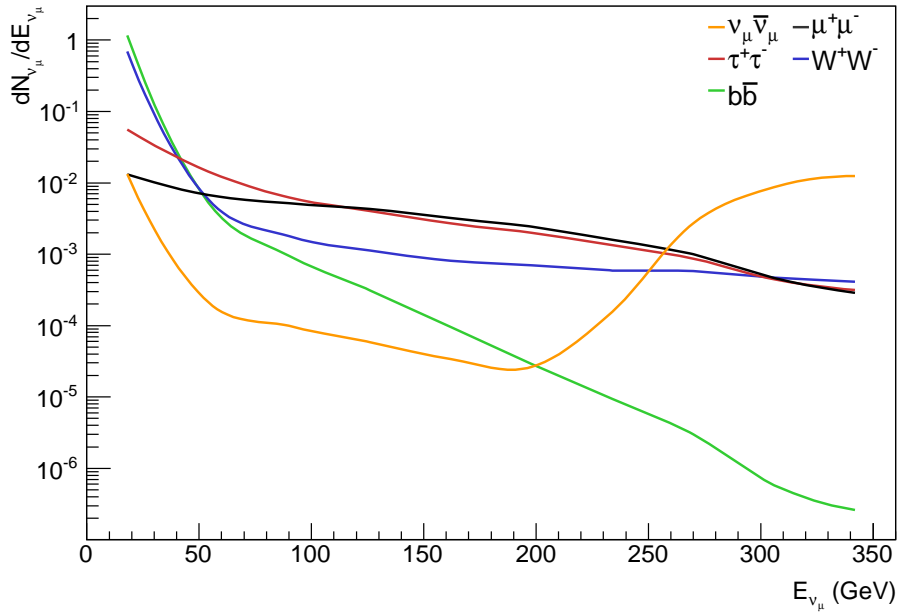}
\caption{Example of muon neutrino energy spectra at the surface of the Earth $\rm dN_{\nu_{\mu}}/dE_{\nu_{\mu}}\Big|_{\oplus}$ as a function of 
their energy $\rm E_{\nu_{\mu}}$ for a WIMP mass $\rm M_{WIMP} = 360$ GeV. The primary self-annihilation channels: 
$\rm WIMP\,WIMP$ $\rm \rightarrow$ $\rm b\bar{b}\,(green)$, $\rm W^{+}W^{-}\,(blue)$, $\rm \tau^{+}\tau^{-}\,(red)$, 
$\rm \mu^{+}\mu^{-}\,(black)$, $\rm \nu_{\mu}\bar{\nu}_{\mu}\,(orange)$ are shown.}
\label{benchchfig}
\end{center}
\end{figure}

\begin{equation}
\rm{\frac{dN_{\nu_{\mu}}}{dE_{\nu_{\mu}}}\Big|_{\oplus} = \sum_{\alpha} P(\nu_{\alpha}\rightarrow\nu_{\mu}) \frac{dN_{\nu_{\alpha}}}{dE_{\nu_{\alpha}}}\Big|_{\ominus}} \, ,
\label{nuoscs}
\end{equation}

\noindent where $\rm dN_{\nu_{\mu}}/dE_{\nu_{\mu}}\Big|_{\oplus}$ is the muon neutrino energy spectrum at the Earth, $\rm P(\nu_{\alpha}\rightarrow\nu_{\mu})$ 
is the probability to observe the oscillation $\rm \nu_{\alpha}\rightarrow\nu_{\mu}$, where $\rm \alpha = e,\mu,\tau$, and $\rm dN_{\nu_{\alpha}}/dE_{\nu_{\alpha}}\Big|_{\ominus}$ 
is the corresponding energy spectrum at the GC. Examples of muon neutrino energy spectra at the surface of the Earth, $\rm dN_{\nu_{\mu}}/dE_{\nu_{\mu}}\Big|_{\oplus}$, 
are shown in Figure~\ref{benchchfig} as a function of their energy $\rm E_{\nu_{\mu}}$ (GeV) (the spectra in muon anti-neutrinos are identical), for an indicative mass 
$\rm M_{WIMP} = 360$ GeV. In this figure, the three-flavour oscillations process in vacuum over the GC-Earth line of sight is used as expressed in Equation~\ref{nuoscs}. 
As with the dark matter search in the direction of the Sun~\cite{dmsun}, $\rm WIMP\,WIMP$$\rm \rightarrow$$\rm b\bar{b}$ is the softest channel of muon neutrino production. The hardest 
channel of muon neutrino production from the GC is the primary $\rm WIMP\,WIMP \rightarrow \nu_{\alpha}\bar{\nu}_{\alpha}$. It is calculated taking into account the contribution of 
electroweak corrections discussed in Ref.~\cite{cirelli}. The three other channels, $\rm WIMP\,WIMP \rightarrow W^{+}W^{-},\,\tau^{+}\tau^{-},\,\mu^{+}\mu^{-}$, present a hard contribution 
to the full spectrum of muon neutrinos at Earth. It can be noted that the channels to $\rm \tau^{+}\tau^{-}$ and $\rm \mu^{+}\mu^{-}$ differ only for $\rm E_{\nu_{\mu}} < 100$ GeV in 
this example. 

In order to be as model-independent as possible, a self-annihilation branching ratio $\rm BR=1$ is used for each of the channels in Expression~\ref{benchch}. Beyond-the-SM particle physics 
with $\rm BR<1$ can be accommodated by scaling the fluxes in Equation~\ref{nuoscs} linearly with the appropriate branching ratio.

\begin{figure}[!t]
\begin{center}
\begin{minipage}[c]{.8\linewidth}
\begin{overpic}[width=\linewidth]{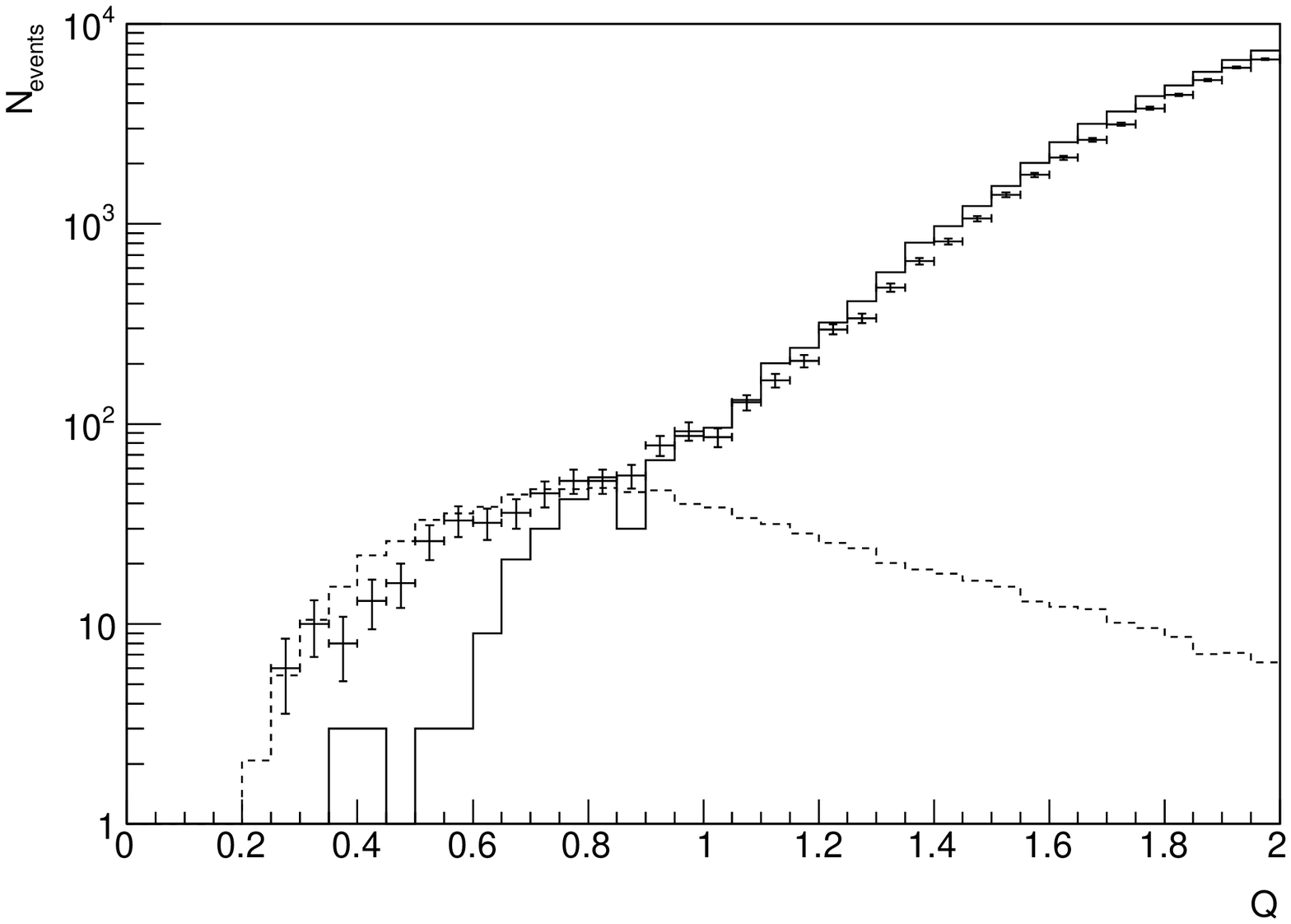}
 \put (15,60) {$\displaystyle QFit$}
\end{overpic}
\begin{overpic}[width=\linewidth]{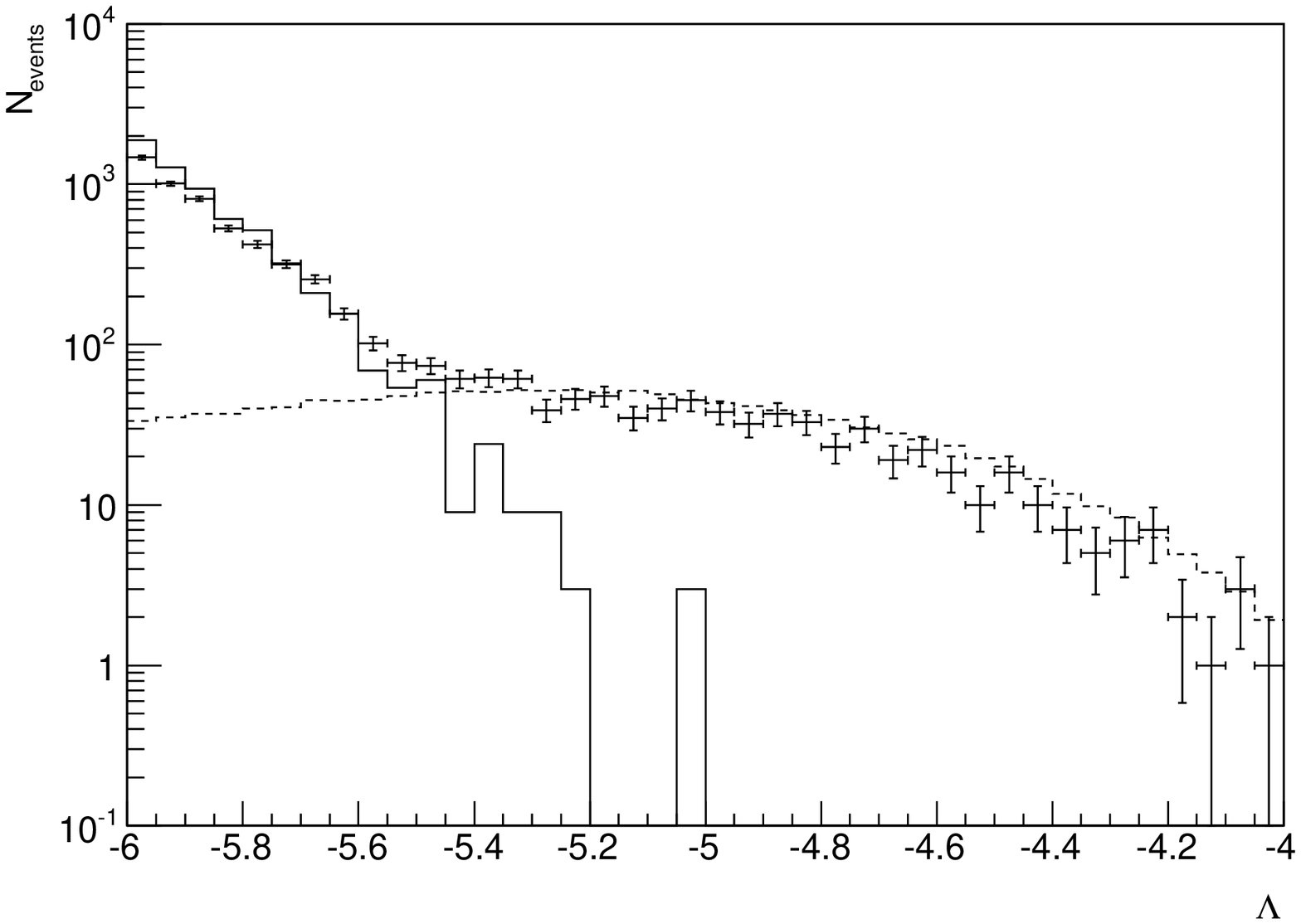}
 \put (85,60) {$\displaystyle\Lambda Fit$}
\end{overpic}
\end{minipage}
\caption{Distribution of the number of reconstructed events, $\rm N_{events}$, for the $\rm QFit$ (top) and $\rm \Lambda Fit$ (bottom) algorithms as a function 
of their respective track reconstruction parameter, $\rm Q$ and $\rm \Lambda$. The expectations according to simulations for atmospheric 
neutrinos (dashed line), atmospheric muons (solid line), and the data (black crosses) for 2012 are presented. For $\rm QFit$, only the upgoing events, 
$\rm \cos(\theta_{rec}) > 0$, are used. For $\rm \Lambda Fit$, the uncertainty of the muon track direction angle is required to be $\rm \beta < 0.5$, 
and the field of view is extended to $\rm \cos(\theta_{rec}) > -0.1$.} 
\label{qcutfig}
\end{center}
\end{figure}

\begin{figure}[!t]
\begin{center}
\begin{minipage}[c]{.8\linewidth}
\begin{overpic}[width=\linewidth]{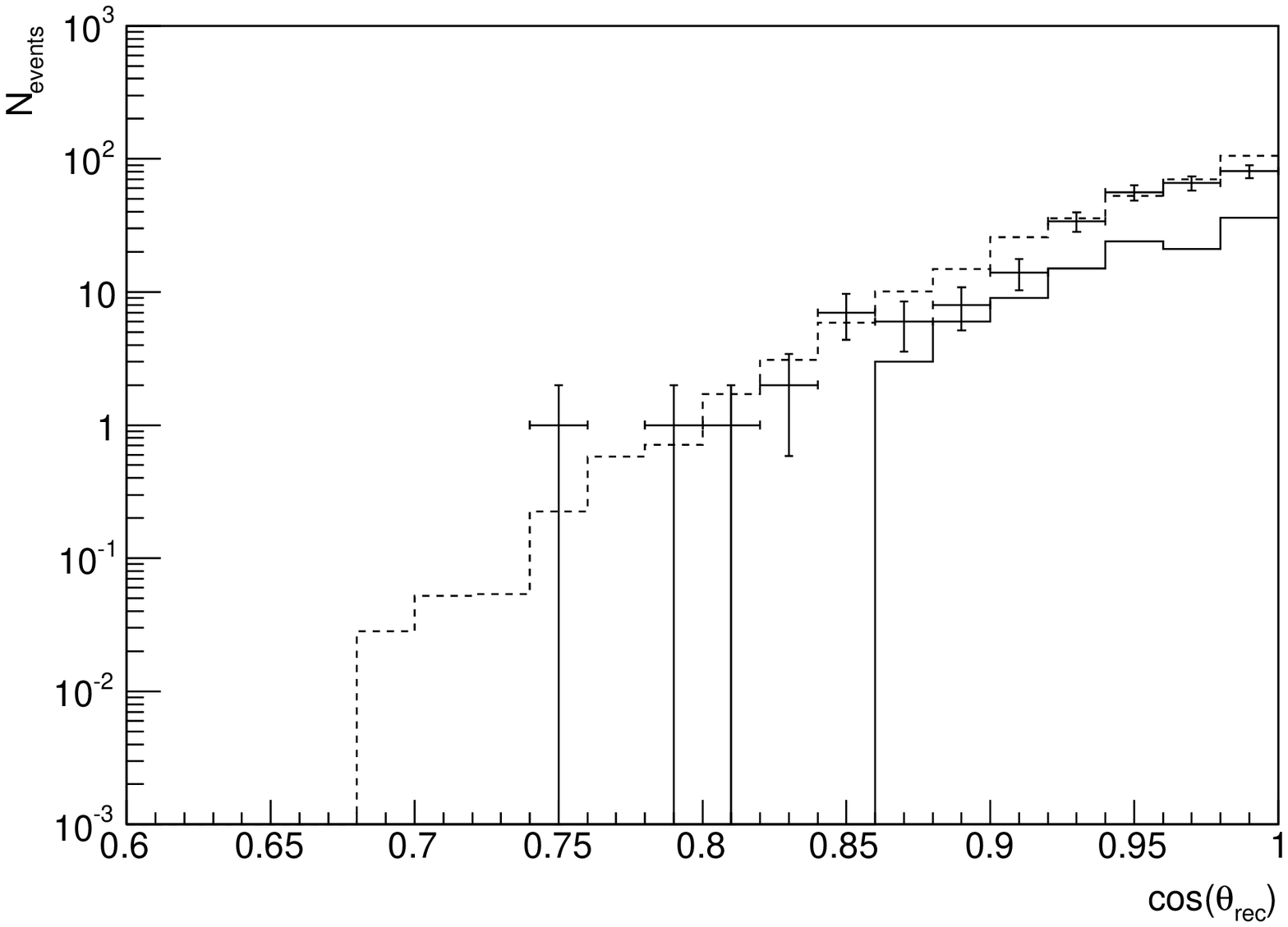}
 \put (15,60) {$\displaystyle QFit$}
\end{overpic}
\begin{overpic}[width=\linewidth]{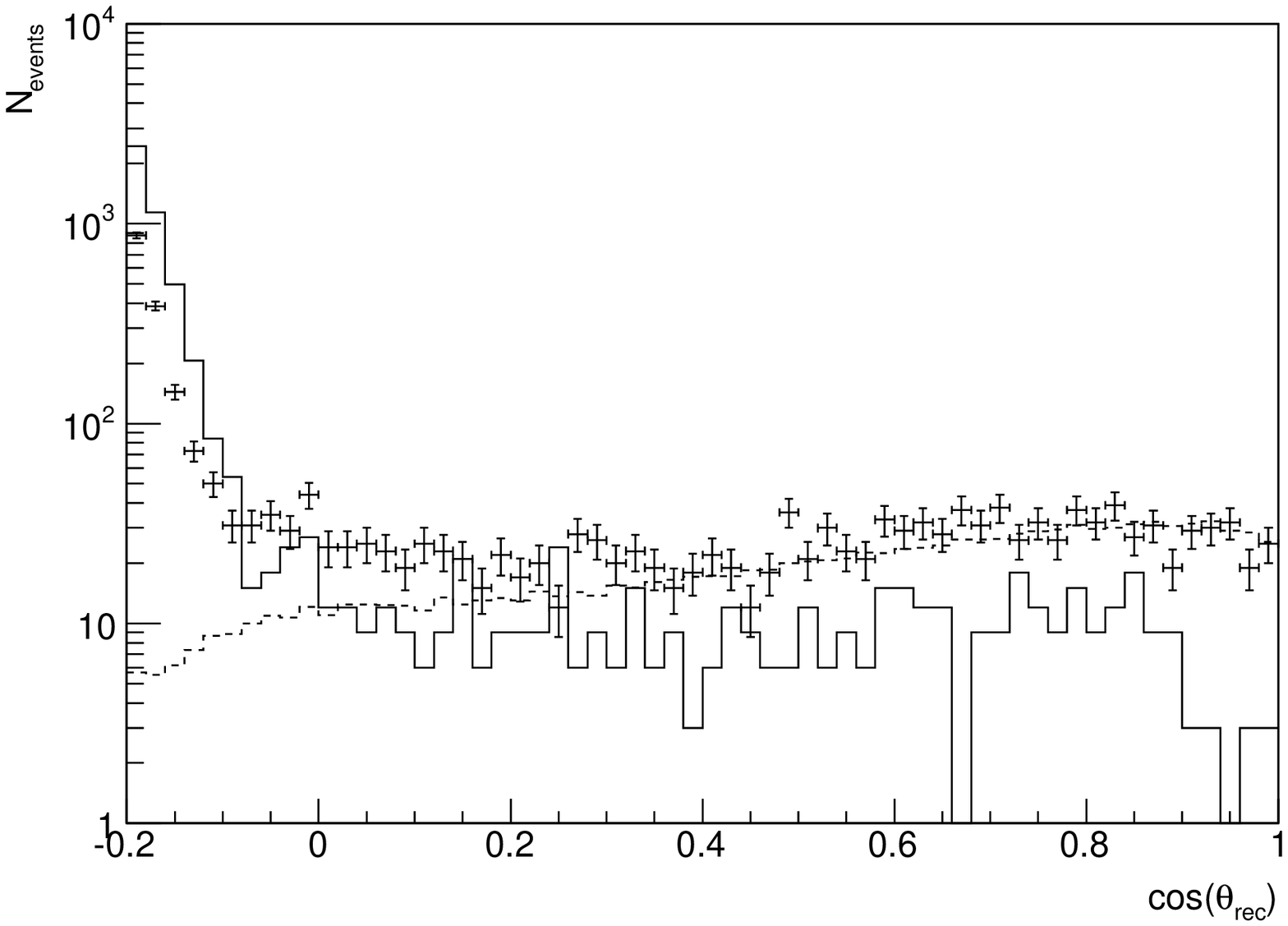}
 \put (85,60) {$\displaystyle\Lambda Fit$}
\end{overpic}
\end{minipage}
\caption{Distribution of the number of reconstructed events, $\rm N_{events}$, for the $\rm QFit$ (top) and $\rm \Lambda Fit$ (bottom) algorithms as a function 
of the cosine of their zenith angle $\rm \cos(\theta_{rec})$. The expectations according to simulations for atmospheric neutrinos (dashed line), atmospheric muons 
(solid line), and the data (black crosses) for 2012 are shown. For $\rm QFit$ and $\rm \Lambda Fit$, the cuts $\rm Q < 0.8$ and ($\rm \Lambda > -5.7$;$\rm \beta < 0.5$) 
are used, respectively.}
\label{costhcutfig}
\end{center}
\end{figure}

\bigskip
The main backgrounds for this analysis are atmospheric muons and neutrinos, both produced in the interactions 
of cosmic rays with the Earth's atmosphere. Downgoing atmospheric muons dominate 
the trigger rate, which ranges from $3$ to $10$ Hz depending on the trigger conditions. They are simulated 
using MUPAGE~\cite{mupage}. Upgoing atmospheric neutrinos, which are recorded at a rate of $\sim$50 $\mu$Hz 
(about four per day), are simulated according to the parameterisation of the atmospheric $\rm \nu_{\mu}$ flux from 
Ref.~\cite{bartol} in the energy range from 10~GeV to 10~PeV using GENHEN~\cite{genhen}. Furthermore, the propagation of 
muon tracks is simulated with the KM3 package~\cite{genhen}. The Cherenkov light produced in the vicinity of the detector 
is propagated taking into account light absorption and scattering in sea water~\cite{transmission}. The characteristics of 
the PMTs are taken from Ref.~\cite{OM} and the overall geometry corresponds to the different layouts of the ANTARES detector during each data-taking period.

The simulated effective area is used to evaluate the expected signal from WIMP self-annihilations. The expected background is estimated from the scrambled data in order to avoid 
systematic uncertainties from the simulation. The scrambling consists in a uniform randomisation of the UTC time of the events in the 
data-taking period. The zenith and azimuth angles of the reconstructed tracks are kept so as to preserve the angular response of 
the detector in the optimisation of the selection criteria. This procedure provides a means to follow a data blinding strategy 
while using all the relevant information on the detector performance.

The criteria to select events and to reduce the background from atmospheric muons and neutrinos, and to improve the 
sensitivity of ANTARES to a dark matter signal, are devised following a blind procedure on the srambled data before 
performing the analysis on the data. 

\section{Optimisation of the event selection criteria}
\label{optimisation}

The data used in this search were recorded between the $\rm 27^{th}$ of January 2007 and the $\rm 31^{st}$ of October 2012, corresponding to a total livetime of about 
$\rm 1321$ days. This livetime is not corrected for the visibility of the GC. During this time, the detector consisted of $5$ lines for most of $\rm 2007$ and $12$ lines 
from $\rm 2008$ to $\rm 2012$, with short periods of $\rm 8, 9$ and 10 lines. 

The $\rm QFit$ and $\rm \Lambda Fit$ methods, as introduced in Section~\ref{antares}, are respectively used for single- and multi-line track fit reconstructions. Pre-selection cuts are applied 
to obtain an event sample, dominated by well-reconstructed atmospheric neutrinos. For $\rm QFit$, only upgoing 
events are used, i.e. $\rm \cos(\theta_{rec}) > 0$. For $\rm \Lambda Fit$, the error estimate of the reconstructed muon track direction, $\beta$, should be smaller 
than $\rm 0.5$, and a cut $\rm \cos(\theta_{rec}) > -0.1$ is applied. The distributions of the track fit quality parameters, $\rm Q$ and $\rm \Lambda$, for the 
resulting event samples are shown in Figure~\ref{qcutfig} for both reconstruction methods. By choosing additionally the cuts $\rm Q < 0.8$ and $\rm \Lambda > -5.7$, 
a purity of $\rm 69$\% and $\rm 72$\% in muon neutrinos is reached for $\rm QFit$ and $\rm \Lambda Fit$, respectively. The zenith angle distributions of the event 
samples after the cuts in $\rm Q$ and $\rm \Lambda$ are shown in Figure~\ref{costhcutfig}. Single-line events, as reconstructed by $\rm QFit$, are mostly found 
close to the vertical direction, as illustrated in Figure~\ref{costhcutfig} (top).

\bigskip
After the pre-selection, the angular separation, $\rm \Psi$, between the reconstructed track and the GC's direction is used in an optimisation of the 
model rejection factor~\cite{mrf} as a function of the WIMP mass, $\rm M_{WIMP}$. For each WIMP mass and annihilation channels listed in the Tables~\ref{tab:qresultsone}+\ref{tab:qresultstwo}+\ref{tab:qresultsthree}+\ref{tab:qresultsfour}, 
the selected value $\rm \Psi$ is the one that minimises the average 90\% confidence level (C.L.) upper limit on the $\rm \nu_{\mu}+\bar{\nu}_{\mu}$ flux, 
$\rm \overline{\Phi}_{\nu_{\mu}+\bar{\nu}_\mu}$, defined as:

\begin{equation}
\rm{\overline{\Phi}_{\nu_{\mu}+\bar{\nu}_\mu} = \frac{\bar{\mu}^{90\%}}{\sum\limits_{i} \bar{A}_{eff}^{i}(M_{WIMP}) \times T_{eff}^{i}}} \, ,
\label{mrfeq}
\end{equation}

\noindent where the index $\rm i$ denotes the periods with different detector configurations, $\rm \bar{\mu}^{90\%}$ is the average upper limit of the 
background from scrambled data at $90$\% C.L. (computed using a Poisson distribution in the Feldman-Cousins approach~\cite{feldmancousins}), and 
$\rm {T_{eff}^{i}}$ is the effective livetime for each detector configuration. The effective area averaged over the neutrino energy, 
$\rm{\bar{A}_{eff}^{i}(M_{\rm WIMP})}$, is defined as:

\begin{eqnarray}
\rm{\bar{A}_{eff}^{i}(M_{\rm WIMP})} & = & 
\rm{\frac{\int_{E_{\nu_{\mu}}^{th}}^{M_{WIMP}} \left ( A_{eff}^{\nu_{\mu},i}+A_{eff}^{\bar{\nu}_{\mu},i} \right ) \, \frac{dN_{\nu_{\mu}}}{dE_{\nu_{\mu}}}\Big|_{\oplus} dE_{\nu_{\mu}}}
{2 \, \int_{0}^{M_{\rm WIMP}}\frac{dN_{\nu_{\mu}}}{dE_{\nu_{\mu}}} dE_{\nu_{\mu}}}} \, ,
\label{aeffeq}
\end{eqnarray}

\begin{figure}[!t]
\begin{center}
\begin{overpic}[scale=0.62]{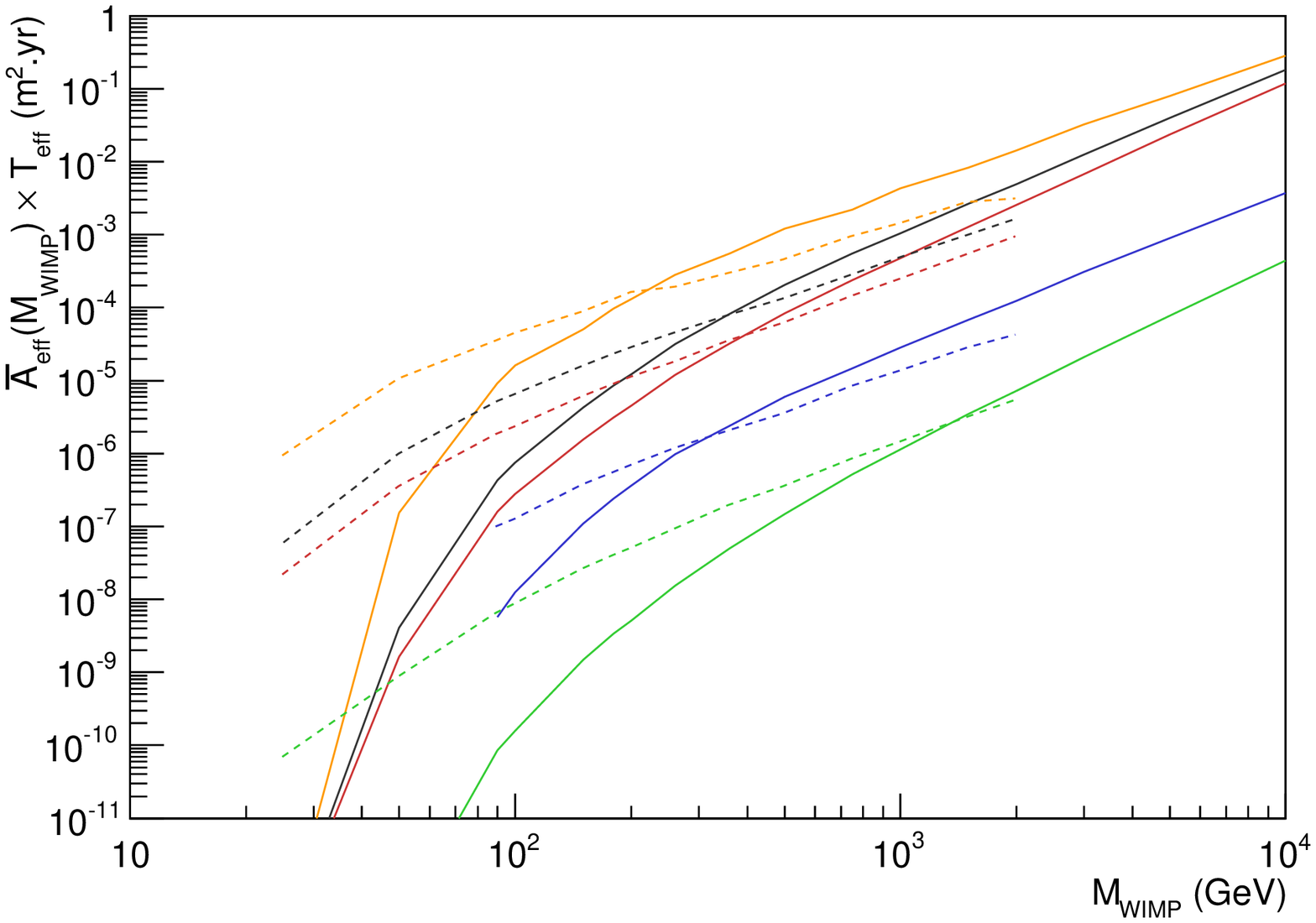}
 \put (70,15){\includegraphics[scale=0.15]{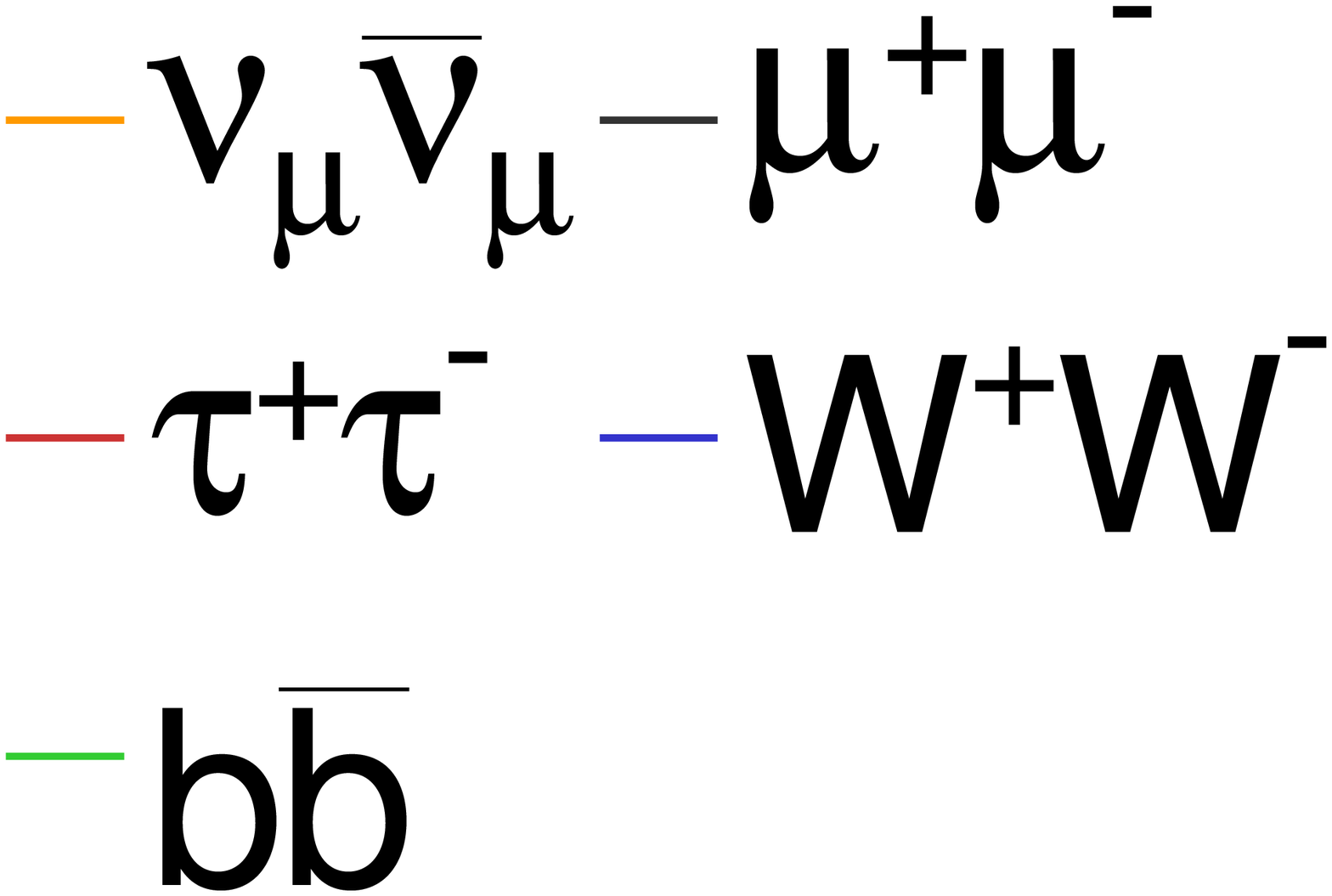}}
 \put (15,55){\includegraphics[scale=0.12]{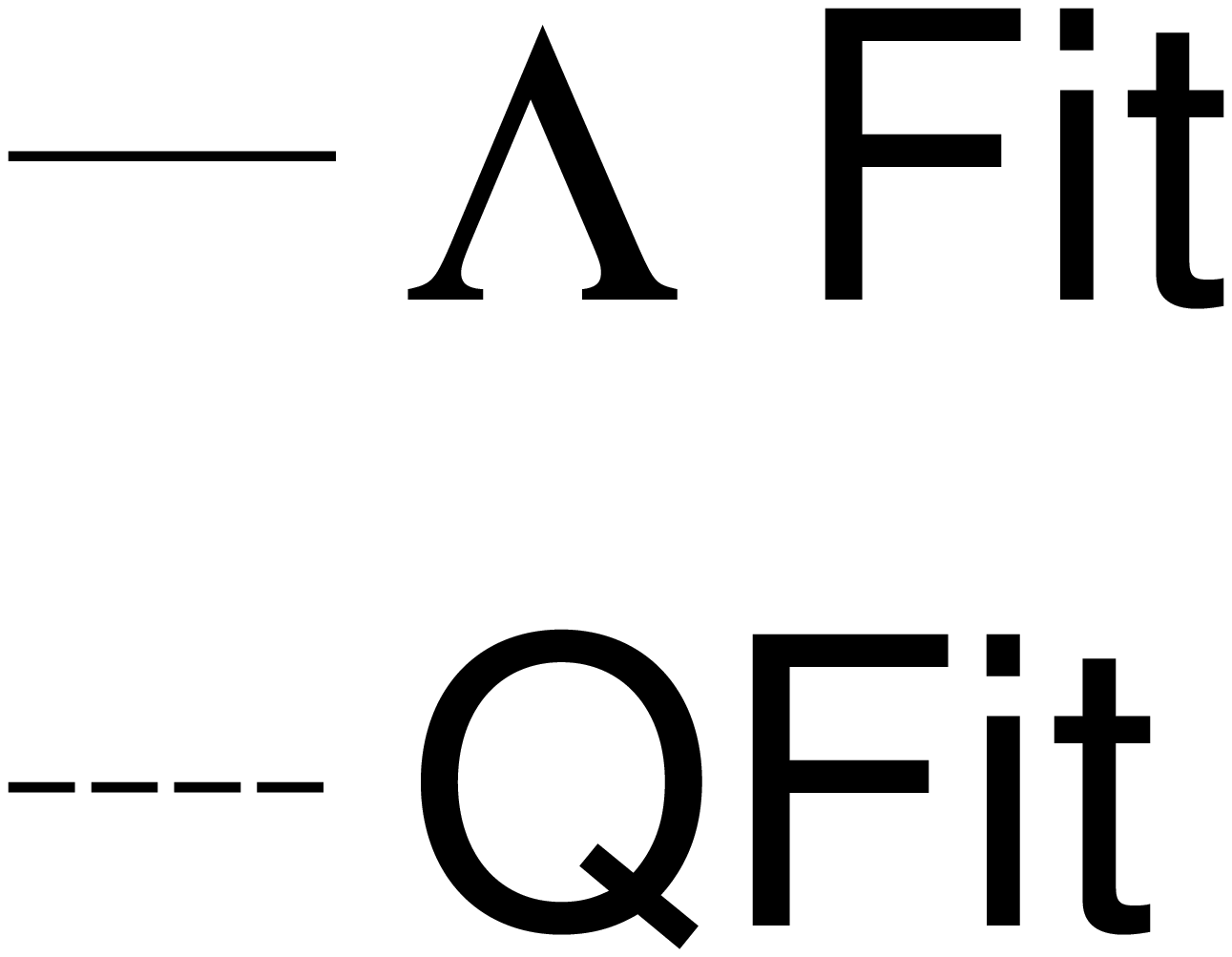}}
\end{overpic}
\caption{Example of the acceptance $\rm \bar{A}_{eff}(M_{\rm WIMP}) \times T_{eff}$ ($\rm m^{2}\cdot yr$) to the signal of WIMP 
self-annihilation towards the GC as a function of WIMP masses $\rm 25\,GeV < M_{WIMP} < 10\,TeV$. The acceptances for the primary self-annihilation 
channels (from bottom to top) $\rm WIMP\,WIMP \rightarrow b\bar{b}(green),W^{+}W^{-} (blue),\tau^{+}\tau^{-} (red),\mu^{+}\mu^{-} (black),\nu_{\mu}\bar{\nu}_{\mu} (orange)$ 
are shown. $\rm QFit$ (dashed lines) and $\rm \Lambda Fit$ (solid lines) are compared.}
\label{aeffmwimpfig}
\end{center}
\end{figure}

\noindent where $\rm E_{\nu_{\mu}}^{th}\simeq15$~GeV is the energy threshold for neutrino detection in ANTARES, 
$\rm dN_{\nu_{\mu}}/dE_{\nu_{\mu}}\Big|_{\oplus}=dN_{\bar{\nu}_{\mu}}/dE_{\bar{\nu}_{\mu}}\Big|_{\oplus}$ is the energy spectrum of 
the neutrinos at the surface of the Earth as shown in Figure~\ref{benchchfig}, and $\rm A_{eff}^{\nu_{\mu}/\bar{\nu}_{\mu}}$ 
is the effective area of ANTARES as a function of the neutrino or anti-neutrino energy for tracks coming from the direction of the GC 
(Figure~\ref{aefffig} for illustration). Due to their different cross-sections, the effective areas for neutrinos and anti-neutrinos are 
slightly different and therefore are considered separately, whereas the fluxes of muon neutrinos and anti-neutrinos from the GC 
are identical. 

The optimisation procedure provides a set of optimum values of angular separation to the GC, $\rm \Psi$, for each 
mass of the WIMP and for each benchmark channel. The distributions of $\rm \Psi$ as a function of the WIMP mass, $\rm M_{WIMP}$, are given in 
Tables~\ref{tab:qresultsone}+\ref{tab:qresultstwo} and~\ref{tab:qresultsthree}+\ref{tab:qresultsfour}. Regardless of self-annihilation channels and reconstruction algorithms, 
an optimised angular separation $\rm \Psi$ to the GC is wider for low $\rm M_{WIMP}$ due to the degradation of the angular resolution $\rm \tilde{\alpha}$ 
at low neutrino energy (Figure~\ref{angresfig}). Moreover, the optimimum $\rm \Psi$ for a given $\rm M_{WIMP}$ depends on the softness of the self-annihilation channel, 
$\rm \Psi$ being wider the softer the channel. 

An example of the acceptance $\rm \bar{A}_{eff}(M_{\rm WIMP}) \times T_{eff}$ is shown in Figure~\ref{aeffmwimpfig}. 
The visibility of the GC and the data-taking periods from 2007 to 2012 are included. The $\rm QFit$ (dashed lines) and 
$\rm \Lambda Fit$ (solid lines) are compared for each self-annihilation channel, inside their respective angular separations, 
$\rm \Psi = 10^{\circ}$ and $\rm \Psi = 2^{\circ}$, that are typical over all the WIMP masses (Tables~\ref{tab:qresultsone}+\ref{tab:qresultstwo}+\ref{tab:qresultsthree}+\ref{tab:qresultsfour}). 
Notice that the softer a self-annihilation channel is, the lower the corresponding acceptance, due to the decrease of the effective area $\rm A_{eff}^{\nu}$ shown in Figure~\ref{aefffig} 
towards the low neutrino energies. Also, the channels $\rm WIMP\,WIMP \rightarrow \nu_{\alpha}\bar{\nu}_{\alpha}$ and 
$\rm WIMP\,WIMP \rightarrow b\bar{b}$ show respectively the largest and smallest acceptance to a WIMP self-annihilation signal. This will 
respectively induce the best and worse limits for a neutrino flux $\rm \Phi_{\nu_{\mu}+\bar{\nu}_\mu}$, as shown in Equation~\ref{mrfeq}. 
The $\rm \bar{A}_{eff}(M_{\rm WIMP}) \times T_{eff}$ distribution of the $W^{+}W^{-}$ channel is kinematically 
allowed for $M_{\rm WIMP} > M_{W} = 80.4$~GeV~\cite{pdg2012}. However, the low mass region, $\rm 25\,GeV < M_{\rm WIMP} < M_{W}$, is probed by 
other channels. Moreover, the energy range for which $\rm QFit$ is more efficient than $\rm \Lambda Fit$ will change slightly according to the 
softness of the self-annihilation channel. As can be seen in Figure~\ref{benchchfig}, most of the contribution of the $b\bar{b}$ channel lies in the 
range $\rm E_{\nu_{\mu}} < 100$ GeV, for $\rm M_{WIMP} = 360$ GeV, where $\rm QFit$ yields a higher effective area compared 
to $\rm \Lambda Fit$. Therefore, $\rm QFit$ is preferred over the range of masses $\rm M_{WIMP} < 1.5$ TeV to compute the best $b\bar{b}$ limit. 
Furthermore, this evolves according to the self-annihilation channel spectrum and the WIMP mass, as described in the following Section~\ref{results}.

\begin{figure}[!t]
\begin{center}
\begin{overpic}[scale=0.62]{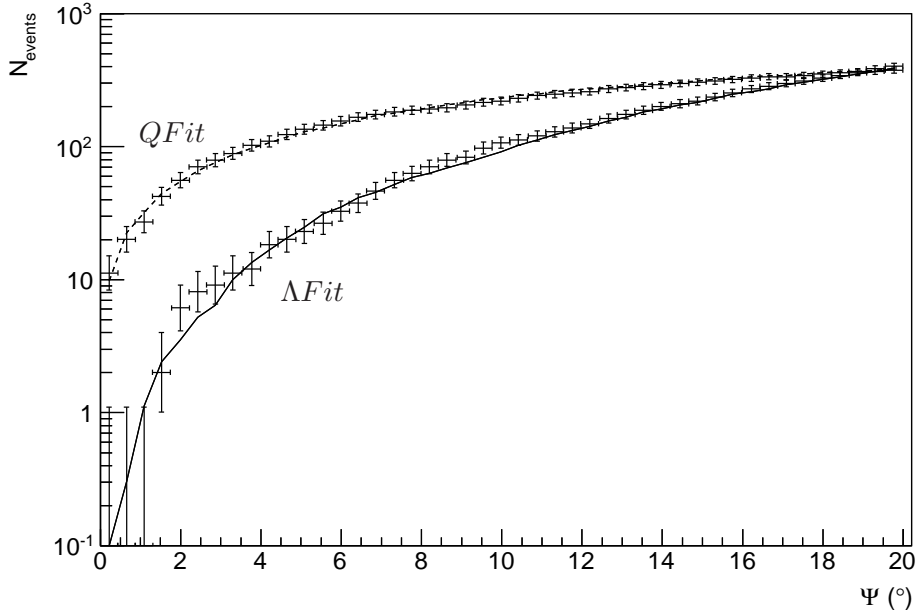}
 \put (15,52) {$\displaystyle QFit$}
 \put (30,35) {$\displaystyle\Lambda Fit$}
\end{overpic}
\end{center}
\caption{Distribution of the number of events as a function of the angular separation, $\rm \Psi$, in the 
direction of the GC for the expected backgrounds (dashed line for $\rm QFit$, solid line for $\rm \Lambda Fit$) compared 
to the data (black crosses) from $\rm QFit$ and $\rm \Lambda Fit$. A $1\sigma$ Gaussian uncertainty is shown for 
each data point.}
\label{databkgfig}
\end{figure}

\begin{figure}[!t]
\begin{center}
\begin{overpic}[scale=0.62]{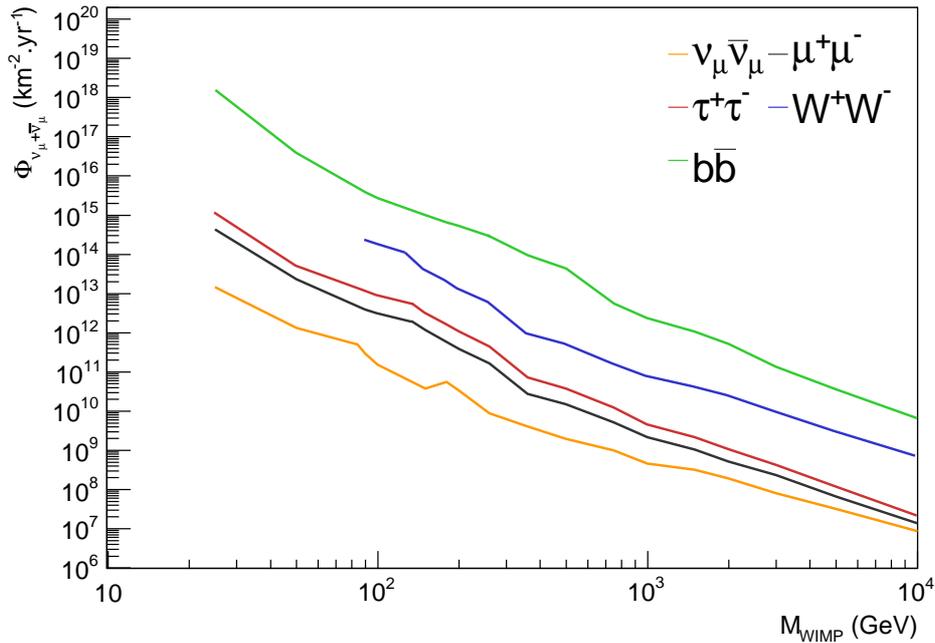}
 \put (70,50){\includegraphics[scale=0.15]{./Channels_Legend.eps}}
\end{overpic}
\caption{$90$\% C.L. upper limits on the neutrino flux, $\rm \Phi_{\nu_{\mu}+\bar{\nu}_\mu}$, 
as a function of the WIMP mass in the range $\rm 25\,GeV < M_{WIMP} < 10\,TeV$ for the self-annihilation channels (from top to bottom) 
$\rm WIMP\,WIMP \rightarrow b\bar{b} (green),W^{+}W^{-} (blue),\tau^{+}\tau^{-} (red),\mu^{+}\mu^{-} (black),\nu_{\mu}\bar{\nu}_{\mu} (orange)$. 
The $\rm QFit$ and $\rm \Lambda Fit$ results are combined.}
\label{phinulimitfig}
\end{center}
\end{figure}

\begin{figure}[!t]
\begin{center}
\begin{overpic}[scale=0.62]{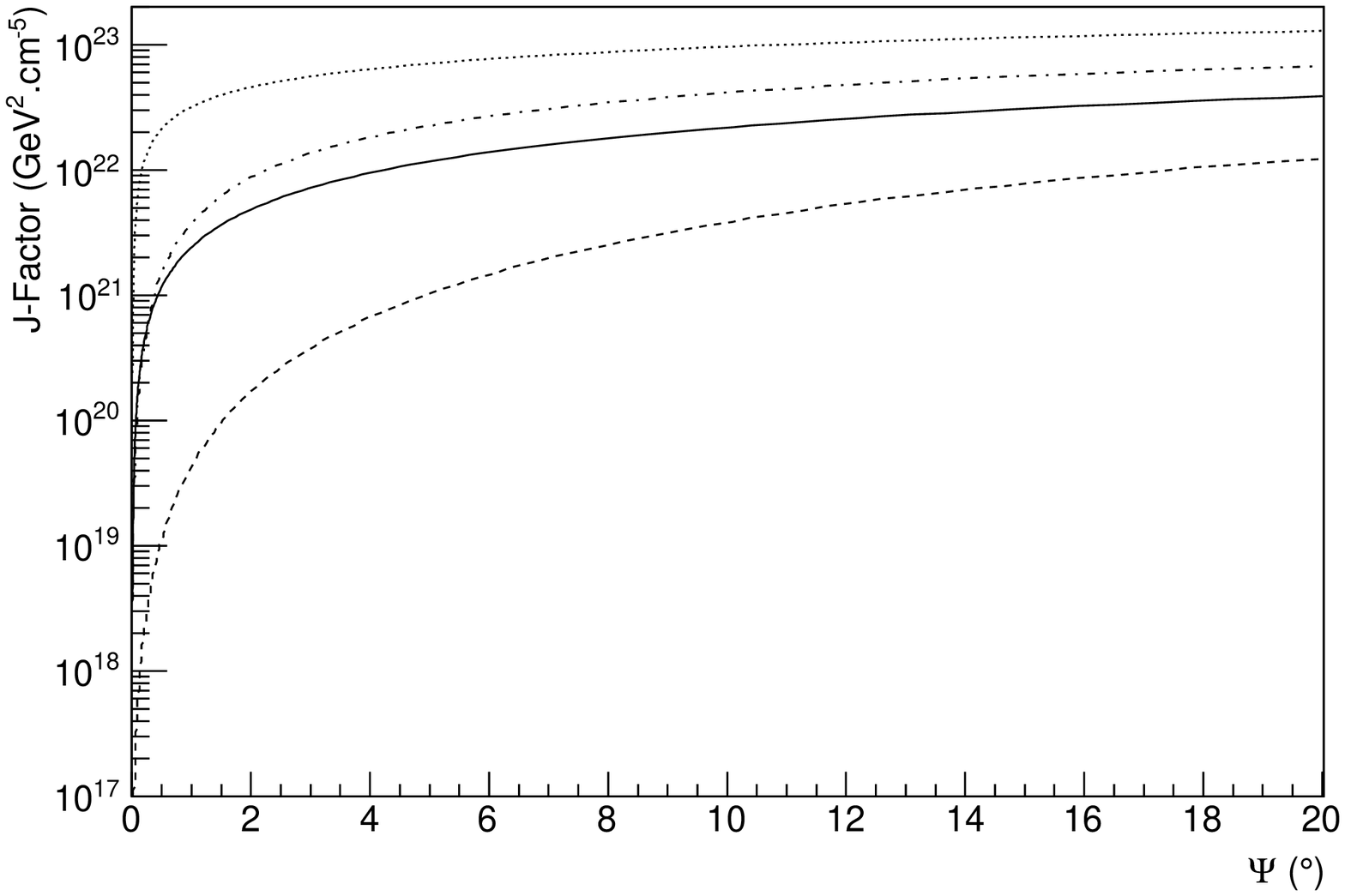}
 \put (60,20){\includegraphics[scale=0.22]{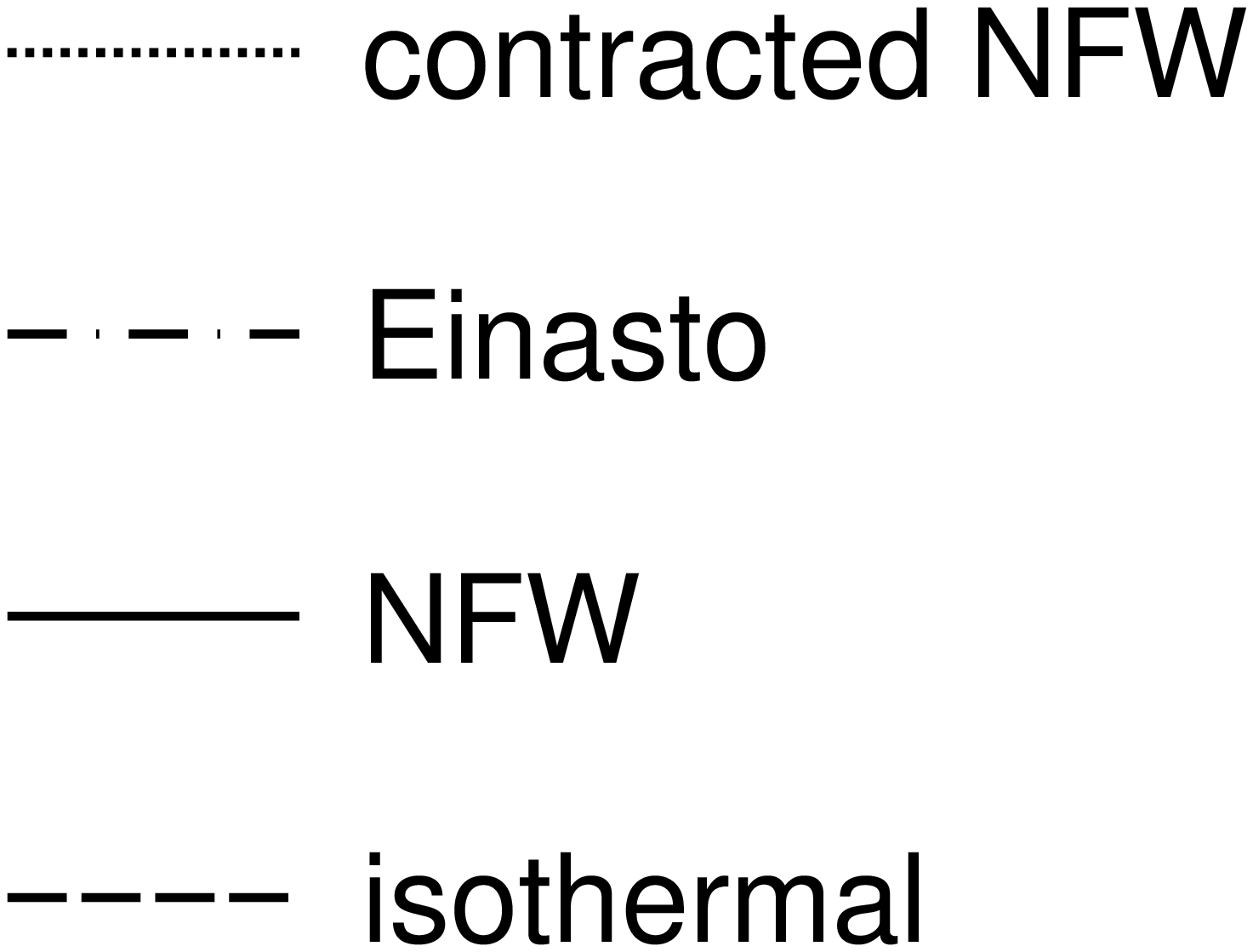}}
\end{overpic}
\caption{Distributions of the integrated J-Factor ($\rm GeV^{2} \cdot cm^{-5}$) as a function of the angular separation to the GC, $\rm \Psi$. 
The dark matter density profiles: a contracted NFW~\cite{contrnfw} (dotted line), Einasto~\cite{einasto} (dot-dashed line), NFW~\cite{nfw} (solid line) 
and isothermal~\cite{isoth} (dashed line), are compared.}
\label{jfactorpsi}
\end{center}
\end{figure}

\begin{figure}[h!]
\begin{center}
\begin{overpic}[scale=0.62]{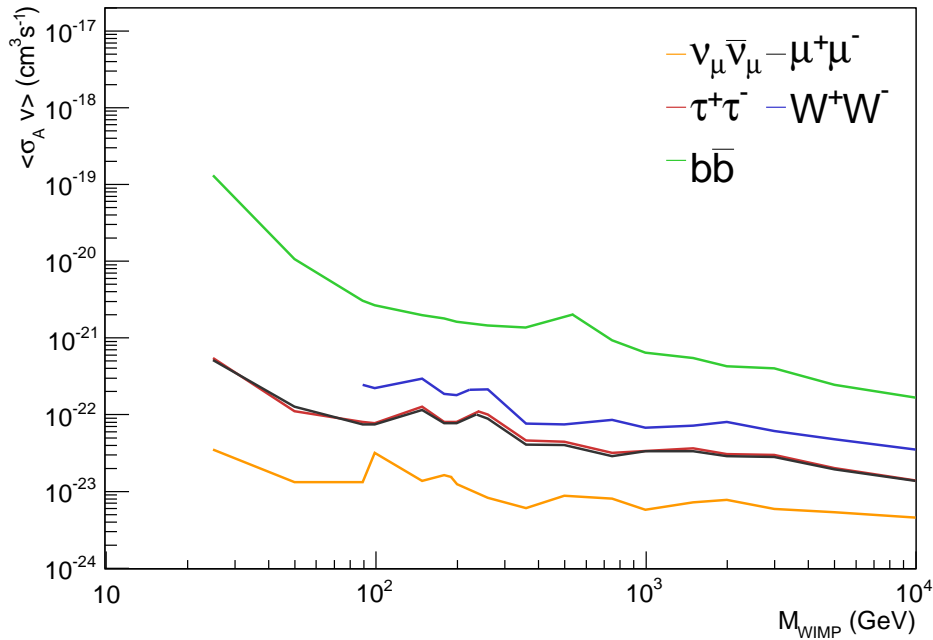}
 \put (70,50){\includegraphics[scale=0.15]{./Channels_Legend.eps}}
\end{overpic}
\caption{The $90$\% C.L. upper limits on the WIMP velocity averaged self-annihilation cross-section, $\rm <\sigma_{A}v>$, 
as a function of the WIMP mass in the range $\rm 25\,GeV < M_{WIMP} < 10\,TeV$ for the self-annihilation channels 
$\rm WIMP\,WIMP \rightarrow b\bar{b} (green)$, $\rm W^{+}W^{-} (blue)$, $\rm \tau^{+}\tau^{-} (red)$, $\rm \mu^{+}\mu^{-} (black)$, $\rm \nu_{\mu}\bar{\nu}_{\mu} (orange)$. 
The $\rm QFit$ and $\rm \Lambda Fit$ results are combined.}
\label{sigmalimitfig}
\end{center}
\end{figure}

\begin{figure}[h!]
\begin{center}
\includegraphics[width=0.8\linewidth]{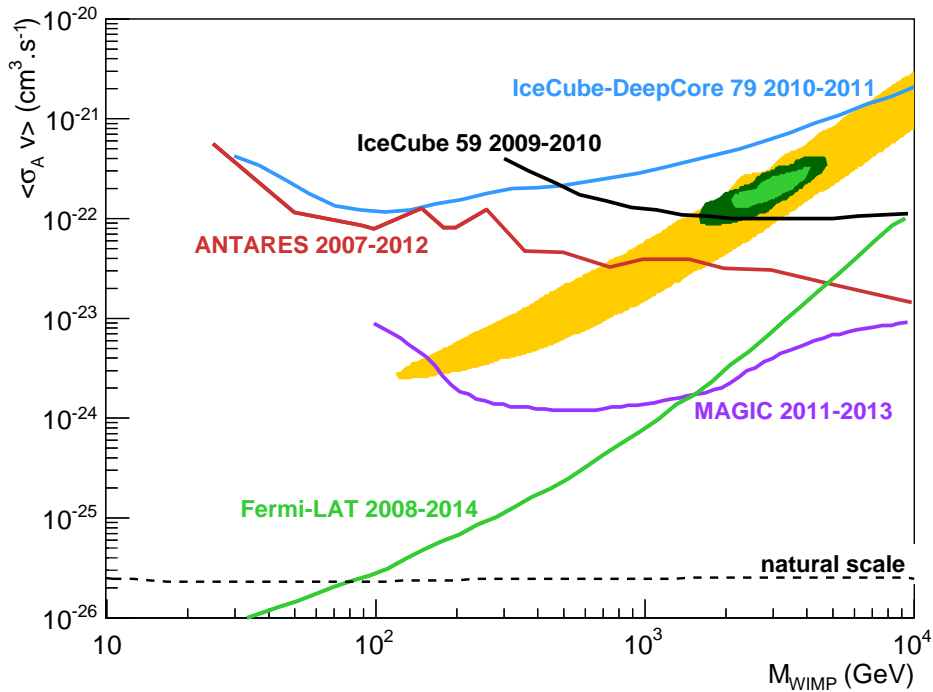}
\caption{The $90$\% C.L. upper limit on the WIMP velocity averaged self-annihilation cross-section, $\rm <\sigma_{A}v>$, 
as a function of the WIMP mass in the range $\rm 25\,GeV < M_{WIMP} < 10\,TeV$ for the self-annihilation channel 
$\rm WIMP\,WIMP \rightarrow \tau^{+}\tau^{-}$ for ANTARES 2007-2012 (red) with $\rm QFit$ and $\rm \Lambda Fit$ results combined. 
This is compared to the limits from IceCube 59 2009-2010~\cite{icecube59} for the Virgo cluster (black), Fermi-LAT 2008-2014~\cite{fermilat} 
for the combined analysis of 15 satellite galaxies (green) and MAGIC 2011-2013~\cite{magic} for Segue 1 (purple), and 
the IceCube-DeepCore 79 2010-2011 sensitivity~\cite{icecube79} for the GC (blue) is also shown. Interpreting observed 
electron/positron excesses as dark matter self-annihilations, the orange (PAMELA) and green (PAMELA, Fermi-LAT and H.E.S.S.) 
ellipses have been obtained~\cite{pamelainterp}. The dashed line indicates the natural scale for which a WIMP is a thermal 
relic of the early Universe~\cite{naturalscale}.}
\label{sigmalimitothersfig}
\end{center}
\end{figure}

\begin{figure}[h!]
\begin{center}
\begin{overpic}[scale=0.62]{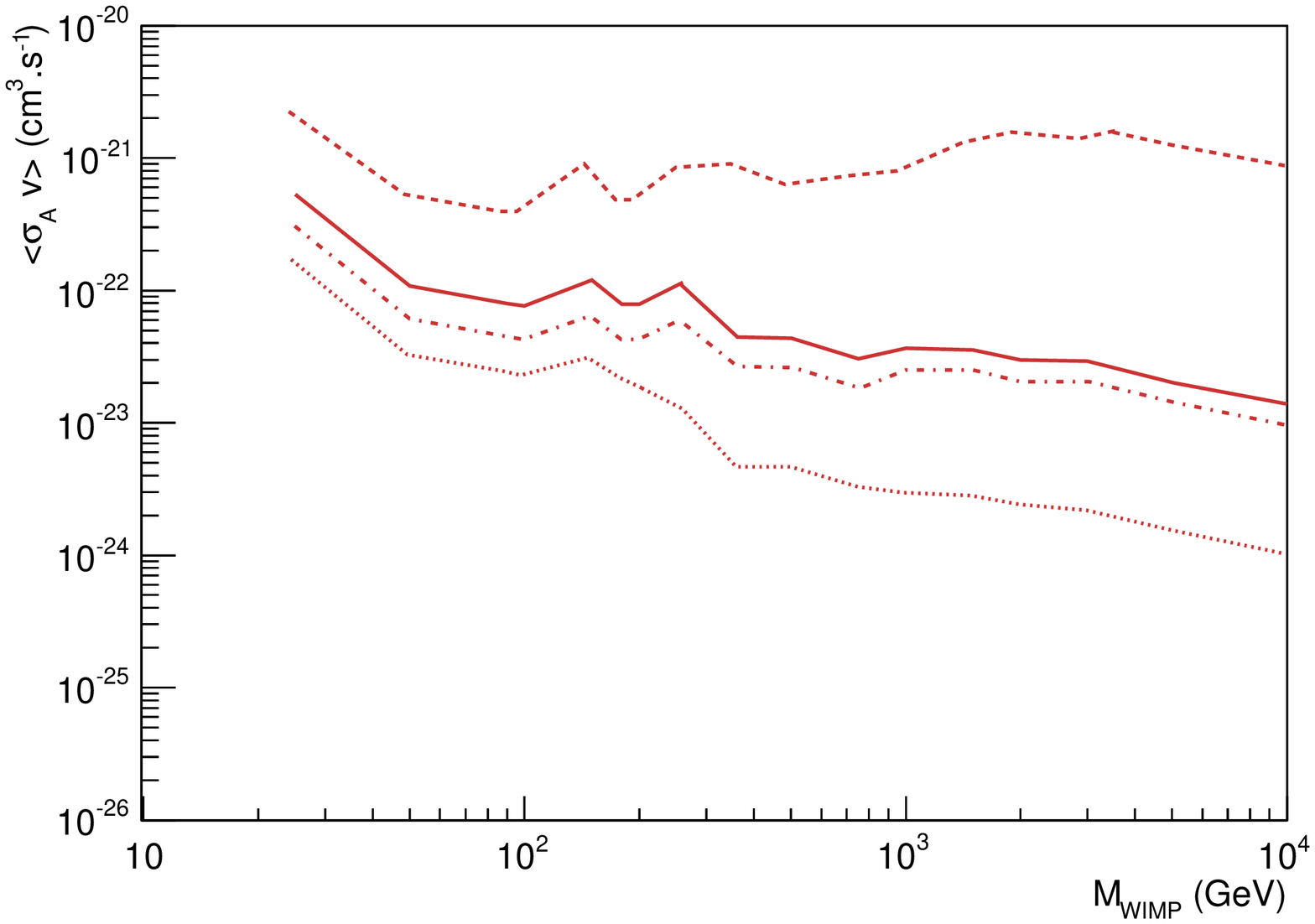}
 \put (15,15){\includegraphics[scale=0.22]{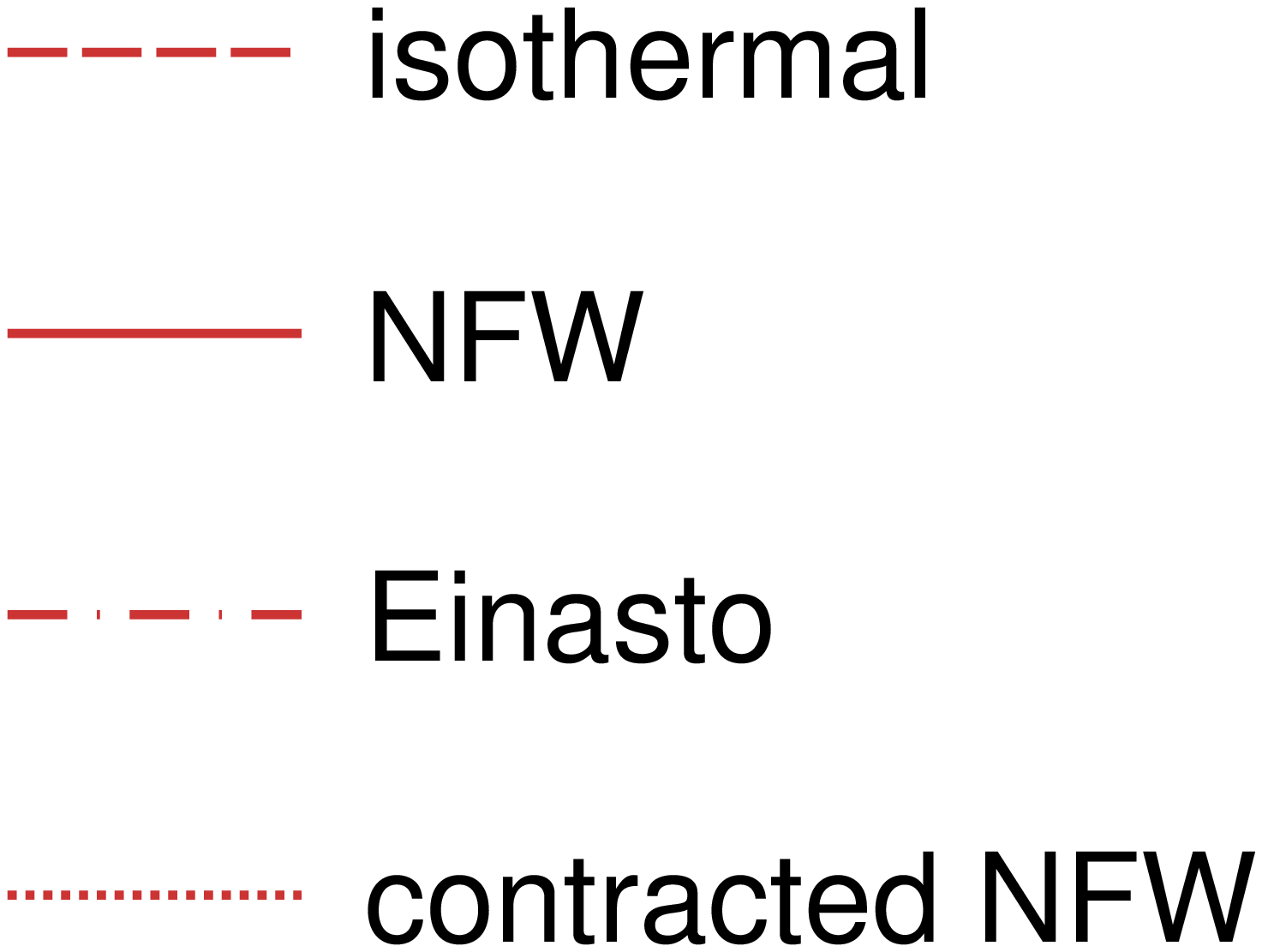}}
\end{overpic}
\caption{The $90$\% C.L. upper limits on the WIMP velocity averaged self-annihilation cross-section, $\rm <\sigma_{A}v>$, 
as a function of the WIMP mass in the range $\rm 25\,GeV < M_{WIMP} < 10\,TeV$ for the self-annihilation channel 
$\rm WIMP\,WIMP \rightarrow \tau^{+}\tau^{-}$. The dark matter density profiles used in the $\rm <\sigma_{A}v>$ computation are NFW~\cite{nfw} (solid line), Einasto~\cite{einasto} (dot-dashed line), 
isothermal~\cite{isoth} (dashed line) and a contracted NFW~\cite{contrnfw} (dotted line).}
\label{sigmalimitsystfig}
\end{center}
\end{figure}

\section{Results and discussion}
\label{results}

Given the pre-selection and optimisation processes described in Section~\ref{antares} and~\ref{optimisation}, 
a total of 369 and 401 events for the unblinded 2007-2012 data are found within an angular separation $\rm \Psi$ of $\rm 20^{\circ}$ 
from $\rm QFit$ and $\rm \Lambda Fit$, respectively. This angle is large enough to contain all values of the optimal angular separation $\rm \Psi$ 
for all WIMP masses and for each channel. Figure~\ref{databkgfig} shows the integrated distribution of the reconstructed event numbers as a function of 
their relative angular separation to the GC, $\rm \Psi$. Data are in good agreement with the expected background for both reconstruction 
algorithms. As no statistically significant excess is observed in the direction of the GC, upper limits on a neutrino flux can be set.

The $90$\% C.L. upper limits on the $\rm \nu_{\mu}+\bar{\nu}_{\mu}$ flux at Earth, $\rm \Phi_{\nu_{\mu}+\bar{\nu}_\mu}$, are computed from the data
according to Equation~\ref{mrfeq}, where the average $90$\% C.L. upper limit $\rm \bar{\mu}^{90\%}$ is replaced by the upper limit at 
$90$\% C.L. on the number of observed events, $\rm \mu^{90\%}$. Systematic uncertainties are taken into account and included in the evaluation of the 
limits using the {\tt Pole} software following the approach described in Ref.~\cite{conrad}. The total systematic uncertainty on the detector efficiency is about 20\% 
and comes mainly from the uncertainties on the average quantum efficiency and the angular acceptance of the PMTs, and the sea water absorption length. 
The detailed uncertainty study is described in Ref.~\cite{timing}. This total systematic uncertainty translates into a degradation of the upper limits between 3\% and 6\%, 
depending on the WIMP mass. The corresponding limits are presented in Figure~\ref{phinulimitfig} for all 
the benchmark self-annihilation channels. In this figure, the results from $\rm QFit$ and $\rm \Lambda Fit$ are combined 
to reach the best upper limit over the whole $\rm M_{WIMP}$ range. Given its soft energy spectrum (Figure~\ref{aeffmwimpfig}), 
the channel $b\bar{b}$ yields the least stringent limit, while it is the opposite for $\nu_{\alpha}\bar{\nu}_{\alpha}$. The upper 
limits, $\rm \Phi_{\nu_{\mu}+\bar{\nu}_\mu}$, improve from low to high WIMP masses, as expected from the general hardness of the neutrino 
spectra at the surface of the Earth, the absence of interaction during the neutrino propagation along the line of sight from the 
GC (no absorption or diffusion are present), and the tightness of the optimum angular separation around the GC, $\rm \Psi$, for high 
WIMP masses (Table~\ref{tab:qresultsfour}). 

\bigskip
The neutrino flux from the self-annihilation of dark matter particles at the GC can be expressed as: 

\begin{eqnarray}
\rm{\Phi_{\nu_{\mu}+\bar{\nu}_\mu}} & = & \rm{\Xi_{\nu_{\mu}+\bar{\nu}_\mu}^{PP} \times J(\Delta\Omega)} \, , \label{phinupsieqone} \\
\rm{\Xi_{\nu_{\mu}+\bar{\nu}_\mu}^{PP}} & = & \rm{\frac{1}{4\pi}\frac{<\sigma_{A}v>}{2M_{WIMP}^{2}} N_{\nu_{\mu}+\bar{\nu}_\mu}} \, , \label{phinupsieqtwo} \\
\rm{J(\Delta\Omega)} & = & \rm{\int_{\Delta\Omega} \int \rho_{DM}^{2}dl\,d\Omega} \, , \label{phinupsieqthree}
\end{eqnarray}

\noindent with $\rm{\Xi_{\nu_{\mu}+\bar{\nu}_\mu}^{PP}}$ as the particle physics term, which depends on the WIMP velocity averaged self-annihilation 
cross-section, $\rm <\sigma_{A}v>$, and the number of neutrinos, $N_{\nu_{\mu}+\bar{\nu}_\mu}$, that reach the surface of the Earth. $N_{\nu_{\mu}+\bar{\nu}_\mu}$ 
is computed from the integration over the neutrino energy of the GC's self-annihilation spectra. The astrophysical J-Factor is the integral over the 
line of sight, $l$, and the solid angle around the GC, $\Omega$, of the dark matter density, $\rm \rho_{DM}$, squared. The density 
$\rm \rho_{DM}$ depends on the chosen dark matter galactic halo profile. The Navarro-Frenk-White (NFW) profile~\cite{nfw} 
is selected as a reference for which the density profile of dark matter is expressed as: 

\begin{equation}
\rm{\rho(r) = \frac{\rho_{s}}{(r/r_{s})(1+r/r_{s})^{2}}} \, ,
\label{nfwprofile}
\end{equation}     

\noindent with $\rm r_{s} = 21.7$ kpc. The normalization of the profile density, $\rm \rho_{s}$, is computed by fixing the dark matter density at the Sun's 
position $\rm \rho(r_{Sun} = 8.5\,kpc) = 0.4\,GeV \cdot cm^{-3}$~\cite{catena, salucci}. The systematic uncertainty introduced by the choice of a 
specific halo profile is discussed at the end of this section.

The J-Factor can be computed using the package CLUMPY~\cite{clumpy} 
with the contribution from clumps turned off. As demonstrated in Equation~\ref{phinupsieqthree}, the J-Factor is a function of the solid angle 
$\rm \Delta\Omega = 2\pi(1-\cos(\Psi))$. Figure~\ref{jfactorpsi} shows the result of the computation of the integrated J-Factor for the NFW profile as a function of $\rm \Psi$. Upper limits for the WIMP velocity-averaged self-annihilation 
cross-section, $\rm <\sigma_{A}v>$, can be set following Equation~\ref{phinupsieqone} for each benchmark self-annihilation channel, with known J-Factor for a given optimum angular separation $\rm \Psi$ for each 
$\rm M_{WIMP}$, and upper limits $\rm \Phi_{\nu_{\mu}+\bar{\nu}_\mu}$ as given in Figure~\ref{phinulimitfig}. Figure~\ref{sigmalimitfig} shows the $90$\% C.L. 
upper limits on $\rm <\sigma_{A}v>$ as a function of the WIMP mass in the range $\rm 25\,GeV < M_{WIMP} < 10\,TeV$ for the whole set of 
self-annihilation channels from Expression~\ref{benchch}. 

The $\rm <\sigma_{A}v>$ upper limits obtained from the 2007-2012 ANTARES 
data are comparable with those obtained by other experiments. Figure~\ref{sigmalimitothersfig} shows the resulting $90$\% C.L. 
upper limit on $\rm <\sigma_{A}v>$ as a function of the WIMP mass in the range $\rm 25\,GeV < M_{WIMP} < 10\,TeV$ for the self-annihilation channel 
$\rm WIMP\,WIMP \rightarrow \tau^{+}\tau^{-}$ from this analysis compared to the results from IceCube-DeepCore 79~\cite{icecube79}, IceCube 59~\cite{icecube59}, 
and to the most stringent gamma-ray limits from Fermi-LAT~\cite{fermilat}, and MAGIC~\cite{magic}. ANTARES 2007-2012 data 
provides the best upper limit at $90$\% C.L. on $\rm <\sigma_{A}v>$ for the channel $\tau^{+}\tau^{-}$ from a neutrino telescope. 
Furthermore, the interpretation~\cite{pamelainterp} of the PAMELA excess as a dark matter self-annihilation signature, after being constrained by Fermi-LAT and H.E.S.S., 
is rejected at $90$\% C.L..

All the results are summarised in Tables~\ref{tab:qresultsone}+\ref{tab:qresultstwo} and~\ref{tab:qresultsthree}+\ref{tab:qresultsfour}, where for each 
WIMP mass and channel the values of the optimised angular separation, $\rm \Psi$, the 90\% C.L. sensitivity, $\rm \overline{\Phi}_{\nu_{\mu}+\bar{\nu}_\mu}$, 
computed from the background without signal expectation, the 90\% C.L. upper limits, $\rm \Phi_{\nu_{\mu}+\bar{\nu}_\mu}$, the acceptance, 
$\rm \bar{A}_{eff}(M_{\rm WIMP}) \times T_{eff}$, and the 90\% C.L. upper limits on $\rm <$$\rm \sigma_{A}v>$ are presented. In these tables, the results shown in 
Figures~\ref{phinulimitfig} and~\ref{sigmalimitfig} are highlighted in bold. To evaluate the influence of the dark matter halo 
profile used in the computation of the $\rm <\sigma_{A}v>$ upper limits, different profiles have been tried. The Einasto profile~\cite{einasto}, 
favoured by recent dark matter-only simulations, is given by: 

\begin{equation}
\rm{\rho(r) = \rho_{s}\exp\{-(2/\alpha)[(r/r_{s})^{\alpha}-1]\}} \, 
\label{einastoprofile}
\end{equation} 

\noindent where $\rm r_{s} = 21.7$ kpc and $\rm \alpha = 0.17$. The isothermal profile~\cite{isoth} 
is given by:

\begin{equation}
\rm{\rho(r) = \frac{\rho_{s}}{1+(r/r_{s})^{2}}} \, 
\label{isoprofile}
\end{equation} 

\noindent where $\rm r_{s} = 4$ kpc. Finally, an adiabatic contraction of the NFW profile due to infall of baryonic matter in the GC region~\cite{contrnfw} is given by: 

\begin{equation}
\rm{\rho(r) = \frac{\rho_{s}}{(r/r_{s})^{\gamma}(1+r/r_{s})^{3-\gamma}}} \, 
\label{contrnfwprofile}
\end{equation} 

\noindent where $\rm \gamma = 1.3$. For all of these profiles, $\rm \rho_{s}$ is computed as for the NFW profile. Figure~\ref{jfactorpsi} 
summarises the different J-Factors for these profiles. Figure~\ref{sigmalimitsystfig} shows the 90\% C.L. upper limit on 
$\rm <\sigma_{A}v>$ for the self-annihilation channel $\rm WIMP\,WIMP \rightarrow \tau^{+}\tau^{-}$ for the different dark matter halo profiles. 
The limits on $\rm <\sigma_{A}v>$ for different profiles vary by one to three orders of magnitude depending on the WIMP mass. The optimum angular 
separation $\rm \Psi$ (Tables~\ref{tab:qresultstwo} and~\ref{tab:qresultsfour}) is usually smaller in the high WIMP mass regime, making the computation 
of the $\rm <\sigma_{A}v>$ more sensitive to the cuspiness of a profile for high WIMP masses (Equation~\ref{phinupsieqthree} and Figure~\ref{jfactorpsi}). 

\section{Summary and conclusion}
\label{conclusion}

Using the 2007-2012 data set recorded by the ANTARES neutrino telescope, an indirect search for dark matter towards the Galactic Centre has been performed. The observed number 
of neutrino events in the Galactic Centre's direction is compatible with the expectation from atmospheric backgrounds (Figure~\ref{databkgfig}), for both 
reconstruction algorithms, optimised for low and high energy events. The 90\% C.L. upper limits have been derived 
for the neutrino flux, $\rm \Phi_{\nu_{\mu}+\bar{\nu}_\mu}$ (Figure~\ref{phinulimitfig}), and the velocity averaged 
self-annihilation cross-section, $\rm <\sigma_{A}v>$ (Figure~\ref{sigmalimitfig}), for all the self-annihilation channels 
$\rm WIMP\,WIMP \rightarrow b\bar{b}$, $\rm W^{+}W^{-}$, $\rm \tau^{+}\tau^{-}$, $\rm \mu^{+}\mu^{-}$, $\rm \nu_{\mu}\bar{\nu}_{\mu}$, in the 
range of WIMP masses $\rm 25\,GeV < M_{WIMP} < 10\,TeV$. The 90\% C.L. upper limit on $\rm <\sigma_{A}v>$ for the channel $\rm WIMP\,WIMP \rightarrow \tau^{+}\tau^{-}$ 
is the most stringent coming from a neutrino telescope, and is complementary to the most constraining upper limits obtained by the Fermi-LAT and MAGIC observatories (Figure~\ref{sigmalimitsystfig}). 
Furthermore, the performance of the ANTARES neutrino telescope through this study allows the rejection at 90\% C.L. of the 
interpretation~\cite{pamelainterp} of the PAMELA electron/positron excess (constrained by Fermi-LAT and H.E.S.S.) as a signal from dark matter 
self-annihilation.   					     

\section*{Acknowledgments}
The authors acknowledge the financial support of the funding agencies:
Centre National de la Recherche Scientifique (CNRS), Commissariat \`a
l'\'ene\-gie atomique et aux \'energies alternatives (CEA), Agence
National de la Recherche (ANR), Commission Europ\'eenne (FEDER fund
and Marie Curie Program), R\'egion Alsace (contrat CPER), R\'egion
Provence-Alpes-C\^ote d'Azur, D\'e\-par\-tement du Var and Ville de La
Seyne-sur-Mer, France; Bundesministerium f\"ur Bildung und Forschung
(BMBF), Germany; Istituto Nazionale di Fisica Nucleare (INFN), Italy;
Stichting voor Fundamenteel Onderzoek der Materie (FOM), Nederlandse
organisatie voor Wetenschappelijk Onderzoek (NWO), the Netherlands;
Council of the President of the Russian Federation for young
scientists and leading scientific schools supporting grants, Russia;
National Authority for Scientific Research (ANCS), Romania; Ministerio
de Ciencia e Innovaci\'on (MICINN), Prometeo of Generalitat Valenciana
and MultiDark, Spain; Agence de l'Oriental and CNRST, Morocco. We also
acknowledge the technical support of Ifremer, AIM and Foselev Marine
for the sea operation and the CC-IN2P3 for the computing facilities.

\newpage

\setlength{\tabcolsep}{1.0pt}

\begin{table*}[!b]
\begin{center}
\begin{footnotesize}
\begin{tabular}{ccccccccccccc}
\hline
\hline
&&&&&&&&&& \\[-0.1in]
$\rm M_{WIMP}$ & Channel & $\rm \Psi$ & $\rm \bar{\mu}^{90\%}$ & $\rm \mu^{90\%}$ & $\rm \bar{A}_{eff}(M_{\rm WIMP})$ & $\rm \overline{\Phi}_{\nu_{\mu}+\bar{\nu}_\mu}$ & $\rm \Phi_{\nu_{\mu}+\bar{\nu}_\mu}$ & $\rm J-Factor$ & $\rm \overline{<\sigma_{A}v>}$ & $\rm <\sigma_{A}v>$ \\
&&&&& $\rm \times T_{eff}$ &&&&& \\
$\rm (GeV)$ & & ($\rm ^{\circ}$) & & & ($\rm m^2.yr$) & ($\rm km^{-2}.yr^{-1}$) & ($\rm km^{-2}.yr^{-1}$) & ($\rm GeV^{2}.cm^{-5}$) & ($\rm cm^{3}.s^{-1}$) & ($\rm cm^{3}.s^{-1}$) \\
&&&&&&&&&& \\[-0.1in]

&&&&&&&&&& \\[-0.1in]
\hline
&&&&&&&&&& \\[-0.1in]
$25$ & $b\bar{b}$ & $14.6$ & $34$ & $22$ & $1.5\times 10^{-11}$ & $2.3\times 10^{18}$ & $\mathbf{1.5\times 10^{18}}$ & $3\times 10^{22}$ & $2\times 10^{-19}$ & $\mathbf{1.3\times 10^{-19}}$ \\
$$ & $\tau\bar{\tau}$ & $14.6$ & $34$ & $22$ & $2.1\times 10^{-8}$ & $1.6\times 10^{15}$ & $\mathbf{1.1\times 10^{15}}$ & $3\times 10^{22}$ & $8.2\times 10^{-22}$ & $\mathbf{5.3\times 10^{-22}}$ \\
$$ & $\mu^{+}\mu^{-}$ & $14.6$ & $34$ & $22$ & $5.5\times 10^{-8}$ & $6.2\times 10^{14}$ & $\mathbf{4\times 10^{14}}$ & $3\times 10^{22}$ & $7.9\times 10^{-22}$ & $\mathbf{5.1\times 10^{-22}}$ \\
$$ & $\nu_{\alpha}\bar{\nu}_{\alpha}$ & $10.3$ & $30$ & $17$ & $1.2\times 10^{-6}$ & $2.5\times 10^{13}$ & $\mathbf{1.4\times 10^{13}}$ & $2.2\times 10^{22}$ & $5.8\times 10^{-23}$ & $\mathbf{3.3\times 10^{-23}}$ \\
&&&&&&&&&& \\[-0.1in]
\hline
&&&&&&&&&& \\[-0.1in]
$50$ & $b\bar{b}$ & $14.6$ & $34$ & $22$ & $6\times 10^{-10}$ & $5.7\times 10^{16}$ & $\mathbf{13.7\times 10^{16}}$ & $3\times 10^{22}$ & $1.6\times 10^{-20}$ & $\mathbf{1\times 10^{-20}}$ \\
$$ & $\tau\bar{\tau}$ & $12.5$ & $32$ & $18$ & $3.8\times 10^{-7}$ & $8.5\times 10^{13}$ & $\mathbf{4.8\times 10^{13}}$ & $2.6\times 10^{22}$ & $1.9\times 10^{-22}$ & $\mathbf{1.1\times 10^{-22}}$ \\
$$ & $\mu^{+}\mu^{-}$ & $12.1$ & $32$ & $21$ & $1\times 10^{-6}$ & $3.1\times 10^{13}$ & $\mathbf{2.1\times 10^{13}}$ & $2.6\times 10^{22}$ & $1.9\times 10^{-22}$ & $\mathbf{1.2\times 10^{-22}}$ \\
$$ & $\nu_{\alpha}\bar{\nu}_{\alpha}$ & $8.95$ & $28$ & $17$ & $1.4\times 10^{-5}$ & $2\times 10^{12}$ & $\mathbf{1.2\times 10^{12}}$ & $2\times 10^{22}$ & $2.1\times 10^{-23}$ & $\mathbf{1.3\times 10^{-23}}$ \\
&&&&&&&&&& \\[-0.1in]
\hline
&&&&&&&&&& \\[-0.1in]
$90$ & $b\bar{b}$ & $12.5$ & $32$ & $18$ & $5\times 10^{-9}$ & $6.4\times 10^{15}$ & $\mathbf{3.6\times 10^{15}}$ & $2.6\times 10^{22}$ & $5.4\times 10^{-21}$ & $\mathbf{3\times 10^{-21}}$ \\
$$ & $W^{+}W^{-}$ & $12$ & $32$ & $20$ & $8.6\times 10^{-8}$ & $3.7\times 10^{14}$ & $\mathbf{2.3\times 10^{14}}$ & $2.5\times 10^{22}$ & $3.7\times 10^{-22}$ & $\mathbf{2.4\times 10^{-22}}$ \\
$$ & $\tau\bar{\tau}$ & $12$ & $32$ & $20$ & $1.9\times 10^{-6}$ & $1.7\times 10^{13}$ & $\mathbf{1\times 10^{13}}$ & $2.5\times 10^{22}$ & $1.3\times 10^{-22}$ & $\mathbf{8\times 10^{-23}}$ \\
$$ & $\mu^{+}\mu^{-}$ & $12$ & $32$ & $20$ & $5.3\times 10^{-6}$ & $6\times 10^{12}$ & $\mathbf{3.8\times 10^{12}}$ & $2.5\times 10^{22}$ & $1.2\times 10^{-22}$ & $\mathbf{7.3\times 10^{-23}}$ \\
$$ & $\nu_{\alpha}\bar{\nu}_{\alpha}$ & $9.95$ & $29$ & $18$ & $4.3\times 10^{-5}$ & $6.8\times 10^{11}$ & $\mathbf{4.2\times 10^{11}}$ & $2.2\times 10^{22}$ & $2.1\times 10^{-23}$ & $\mathbf{1.3\times 10^{-23}}$ \\
&&&&&&&&&& \\[-0.1in]
\hline
&&&&&&&&&& \\[-0.1in]
$100$ & $b\bar{b}$ & $12.5$ & $32$ & $18$ & $7\times 10^{-9}$ & $4.6\times 10^{15}$ & $\mathbf{2.6\times 10^{15}}$ & $2.6\times 10^{22}$ & $4.6\times 10^{-21}$ & $\mathbf{2.6\times 10^{-21}}$ \\
$$ & $W^{+}W^{-}$ & $12$ & $32$ & $20$ & $1.2\times 10^{-7}$ & $2.7\times 10^{14}$ & $\mathbf{1.7\times 10^{14}}$ & $2.5\times 10^{22}$ & $3.4\times 10^{-22}$ & $\mathbf{2.1\times 10^{-22}}$ \\
$$ & $\tau\bar{\tau}$ & $12$ & $32$ & $20$ & $2.5\times 10^{-6}$ & $1.3\times 10^{13}$ & $\mathbf{8.1\times 10^{12}}$ & $2.5\times 10^{22}$ & $1.2\times 10^{-22}$ & $\mathbf{7.7\times 10^{-23}}$ \\
$$ & $\mu^{+}\mu^{-}$ & $12$ & $32$ & $20$ & $6.8\times 10^{-6}$ & $4.7\times 10^{12}$ & $\mathbf{2.9\times 10^{12}}$ & $2.5\times 10^{22}$ & $1.1\times 10^{-22}$ & $\mathbf{7.1\times 10^{-23}}$ \\
$$ & $\nu_{\alpha}\bar{\nu}_{\alpha}$ & $6.95$ & $25$ & $29$ & $4.7\times 10^{-5}$ & $5.3\times 10^{11}$ & $6.1\times 10^{11}$ & $1.6\times 10^{22}$ & $2.7\times 10^{-23}$ & $3.2\times 10^{-23}$ \\
&&&&&&&&&& \\[-0.1in]
\hline
&&&&&&&&&& \\[-0.1in]
$150$ & $b\bar{b}$ & $12$ & $32$ & $20$ & $2.1\times 10^{-8}$ & $1.5\times 10^{15}$ & $\mathbf{9.6\times 10^{14}}$ & $2.5\times 10^{22}$ & $3.1\times 10^{-21}$ & $\mathbf{1.9\times 10^{-21}}$ \\
$$ & $W^{+}W^{-}$ & $8.15$ & $27$ & $21$ & $2.8\times 10^{-7}$ & $9.5\times 10^{13}$ & $7.4\times 10^{13}$ & $1.8\times 10^{22}$ & $3.7\times 10^{-22}$ & $2.9\times 10^{-22}$ \\
$$ & $\tau\bar{\tau}$ & $8.15$ & $27$ & $21$ & $5.1\times 10^{-6}$ & $5.3\times 10^{12}$ & $\mathbf{4.1\times 10^{12}}$ & $1.8\times 10^{22}$ & $1.5\times 10^{-22}$ & $\mathbf{1.2\times 10^{-22}}$ \\
$$ & $\mu^{+}\mu^{-}$ & $8.15$ & $27$ & $21$ & $1.4\times 10^{-5}$ & $2\times 10^{12}$ & $\mathbf{1.5\times 10^{12}}$ & $1.8\times 10^{22}$ & $1.5\times 10^{-22}$ & $\mathbf{1.1\times 10^{-22}}$ \\
$$ & $\nu_{\alpha}\bar{\nu}_{\alpha}$ & $9.75$ & $29$ & $18$ & $1.1\times 10^{-4}$ & $2.7\times 10^{11}$ & $1.7\times 10^{11}$ & $2.1\times 10^{22}$ & $2.2\times 10^{-23}$ & $1.4\times 10^{-23}$ \\
&&&&&&&&&& \\[-0.1in]
\hline
&&&&&&&&&& \\[-0.1in]
$180$ & $b\bar{b}$ & $12$ & $32$ & $20$ & $3.2\times 10^{-8}$ & $1\times 10^{15}$ & $\mathbf{6.3\times 10^{14}}$ & $2.5\times 10^{22}$ & $2.7\times 10^{-21}$ & $\mathbf{1.7\times 10^{-21}}$ \\
$$ & $W^{+}W^{-}$ & $9.75$ & $29$ & $18$ & $4.6\times 10^{-7}$ & $6.3\times 10^{13}$ & $3.9\times 10^{13}$ & $2.1\times 10^{22}$ & $3\times 10^{-22}$ & $1.8\times 10^{-22}$ \\
$$ & $\tau\bar{\tau}$ & $9.75$ & $29$ & $18$ & $8.1\times 10^{-6}$ & $3.6\times 10^{12}$ & $2.2\times 10^{12}$ & $2.1\times 10^{22}$ & $1.3\times 10^{-22}$ & $7.9\times 10^{-23}$ \\
$$ & $\mu^{+}\mu^{-}$ & $9.75$ & $29$ & $18$ & $2.1\times 10^{-5}$ & $1.4\times 10^{12}$ & $8.5\times 10^{11}$ & $2.1\times 10^{22}$ & $1.2\times 10^{-22}$ & $7.5\times 10^{-23}$ \\
$$ & $\nu_{\alpha}\bar{\nu}_{\alpha}$ & $9.15$ & $28$ & $19$ & $1.4\times 10^{-4}$ & $2.1\times 10^{11}$ & $1.4\times 10^{11}$ & $2\times 10^{22}$ & $2.4\times 10^{-23}$ & $1.6\times 10^{-23}$ \\
&&&&&&&&&& \\[-0.1in]
\hline
&&&&&&&&&& \\[-0.1in]
$200$ & $b\bar{b}$ & $12$ & $32$ & $20$ & $4.1\times 10^{-8}$ & $7.8\times 10^{14}$ & $\mathbf{4.9\times 10^{14}}$ & $2.5\times 10^{22}$ & $2.5\times 10^{-21}$ & $\mathbf{1.6\times 10^{-21}}$ \\
$$ & $W^{+}W^{-}$ & $9.75$ & $29$ & $18$ & $5.9\times 10^{-7}$ & $4.9\times 10^{13}$ & $3\times 10^{13}$ & $2.1\times 10^{22}$ & $2.8\times 10^{-22}$ & $1.7\times 10^{-22}$ \\
$$ & $\tau\bar{\tau}$ & $9.75$ & $29$ & $18$ & $1\times 10^{-5}$ & $2.9\times 10^{12}$ & $1.8\times 10^{12}$ & $2.1\times 10^{22}$ & $1.3\times 10^{-22}$ & $7.9\times 10^{-23}$ \\
$$ & $\mu^{+}\mu^{-}$ & $9.75$ & $29$ & $18$ & $2.6\times 10^{-5}$ & $1.1\times 10^{12}$ & $7\times 10^{11}$ & $2.1\times 10^{22}$ & $1.2\times 10^{-22}$ & $7.5\times 10^{-23}$ \\
$$ & $\nu_{\alpha}\bar{\nu}_{\alpha}$ & $9.15$ & $28$ & $19$ & $1.7\times 10^{-4}$ & $1.6\times 10^{11}$ & $1.1\times 10^{11}$ & $2\times 10^{22}$ & $2.2\times 10^{-23}$ & $1.5\times 10^{-23}$ \\
&&&&&&&&&& \\[-0.1in]
\hline
&&&&&&&&&& \\[-0.1in]
$260$ & $b\bar{b}$ & $12$ & $32$ & $20$ & $7.2\times 10^{-8}$ & $4.4\times 10^{14}$ & $\mathbf{2.8\times 10^{14}}$ & $2.5\times 10^{22}$ & $2.2\times 10^{-21}$ & $\mathbf{1.4\times 10^{-21}}$ \\
$$ & $W^{+}W^{-}$ & $7.85$ & $27$ & $21$ & $9\times 10^{-7}$ & $3\times 10^{13}$ & $2.3\times 10^{13}$ & $1.8\times 10^{22}$ & $3.4\times 10^{-22}$ & $2.7\times 10^{-22}$ \\
$$ & $\tau\bar{\tau}$ & $8.05$ & $27$ & $20$ & $1.5\times 10^{-5}$ & $1.8\times 10^{12}$ & $1.3\times 10^{12}$ & $1.8\times 10^{22}$ & $1.5\times 10^{-22}$ & $1.1\times 10^{-22}$ \\
$$ & $\mu^{+}\mu^{-}$ & $8.05$ & $27$ & $20$ & $3.7\times 10^{-5}$ & $7.2\times 10^{11}$ & $5.4\times 10^{11}$ & $1.8\times 10^{22}$ & $1.5\times 10^{-22}$ & $1.1\times 10^{-22}$ \\
$$ & $\nu_{\alpha}\bar{\nu}_{\alpha}$ & $7.85$ & $27$ & $21$ & $2\times 10^{-4}$ & $1.3\times 10^{11}$ & $1\times 10^{11}$ & $1.8\times 10^{22}$ & $3.1\times 10^{-23}$ & $2.4\times 10^{-23}$ \\
&&&&&&&&&& \\[-0.1in]
\hline
&&&&&&&&&& \\[-0.1in]
$360$ & $b\bar{b}$ & $10.5$ & $30$ & $18$ & $1.3\times 10^{-7}$ & $2.3\times 10^{14}$ & $1.4\times 10^{14}$ & $2.3\times 10^{22}$ & $2.2\times 10^{-21}$ & $1.3\times 10^{-21}$ \\
$$ & $W^{+}W^{-}$ & $9.25$ & $29$ & $17$ & $1.6\times 10^{-6}$ & $1.7\times 10^{13}$ & $1\times 10^{13}$ & $2\times 10^{22}$ & $3.1\times 10^{-22}$ & $1.9\times 10^{-22}$ \\
$$ & $\tau\bar{\tau}$ & $8.05$ & $27$ & $20$ & $2.6\times 10^{-5}$ & $1\times 10^{12}$ & $7.6\times 10^{11}$ & $1.8\times 10^{22}$ & $1.6\times 10^{-22}$ & $1.2\times 10^{-22}$ \\
$$ & $\mu^{+}\mu^{-}$ & $8.05$ & $27$ & $20$ & $6.2\times 10^{-5}$ & $4.4\times 10^{11}$ & $3.2\times 10^{11}$ & $1.8\times 10^{22}$ & $1.6\times 10^{-22}$ & $1.2\times 10^{-22}$ \\
$$ & $\nu_{\alpha}\bar{\nu}_{\alpha}$ & $9.35$ & $29$ & $16$ & $3.1\times 10^{-4}$ & $9.2\times 10^{10}$ & $5.1\times 10^{10}$ & $2.1\times 10^{22}$ & $3.1\times 10^{-23}$ & $1.7\times 10^{-23}$ \\

\hline
\hline
\end{tabular}
\caption{Results after optimisation from $\rm QFit$ for the angular separation, $\Psi$, the average $90\%$ C.L. upper limit on the expected signal, $\rm \bar{\mu}^{90\%}$; the $90\%$ C.L. upper limit on the expected signal, 
$\rm \mu^{90\%}$; the total acceptance, $\rm \bar{A}_{eff}(M_{WIMP})\times T_{eff}$; the $90\%$ C.L. sensitivity on the neutrino flux at Earth, $\rm \overline{\Phi}_{\nu_{\mu}+\bar{\nu}_\mu}$; 
the $90\%$ C.L. upper limit on the neutrino flux at Earth, $\rm \Phi_{\nu_{\mu}+\bar{\nu}_\mu}$; the J-Factor for the given $\Psi$; the $90\%$ C.L. sensitivity on the velocity averaged 
self-annihilation cross-section, $\rm \overline{<\sigma_{A}v>}$; and the corresponding $90\%$ C.L. upper limit on $\rm <\sigma_{A}v>$. Results for $\rm M_{\rm WIMP}>360$ GeV are available in 
Table~\ref{tab:qresultstwo}. The results shown in Figures~\ref{phinulimitfig} and~\ref{sigmalimitfig} are indicated in bold characters.
\label{tab:qresultsone}}
\end{footnotesize}
\end{center}
\end{table*}

\begin{table*}[!b]
\begin{center}
\begin{footnotesize}
\begin{tabular}{ccccccccccc}
\hline
\hline
&&&&&&&&&& \\[-0.1in]
$\rm M_{WIMP}$ & Channel & $\rm \Psi$ & $\rm \bar{\mu}^{90\%}$ & $\rm \mu^{90\%}$ & $\rm \bar{A}_{eff}(M_{\rm WIMP})$ & $\rm \overline{\Phi}_{\nu_{\mu}+\bar{\nu}_\mu}$ & $\rm \Phi_{\nu_{\mu}+\bar{\nu}_\mu}$ & $\rm J-Factor$ & $\rm \overline{<\sigma_{A}v>}$ & $\rm <\sigma_{A}v>$ \\
&&&&& $\rm \times T_{eff}$ &&&&& \\
$\rm (GeV)$ & & ($\rm ^{\circ}$) & & & ($\rm m^2.yr$) & ($\rm km^{-2}.yr^{-1}$) & ($\rm km^{-2}.yr^{-1}$) & ($\rm GeV^{2}.cm^{-5}$) & ($\rm cm^{3}.s^{-1}$) & ($\rm cm^{3}.s^{-1}$) \\
&&&&&&&&&& \\[-0.1in]

&&&&&&&&&& \\[-0.1in]
\hline
&&&&&&&&&& \\[-0.1in]
$500$ & $b\bar{b}$ & $8.15$ & $27$ & $21$ & $2.2\times 10^{-7}$ & $1.2\times 10^{14}$ & $9.6\times 10^{13}$ & $1.8\times 10^{22}$ & $2.5\times 10^{-21}$ & $2\times 10^{-21}$ \\
$$ & $W^{+}W^{-}$ & $9.25$ & $29$ & $17$ & $2.6\times 10^{-6}$ & $1.1\times 10^{13}$ & $6.4\times 10^{12}$ & $2\times 10^{22}$ & $3.5\times 10^{-22}$ & $2.1\times 10^{-22}$ \\
$$ & $\tau\bar{\tau}$ & $9.25$ & $29$ & $17$ & $4.7\times 10^{-5}$ & $6\times 10^{11}$ & $3.6\times 10^{11}$ & $2\times 10^{22}$ & $1.6\times 10^{-22}$ & $9.7\times 10^{-23}$ \\
$$ & $\mu^{+}\mu^{-}$ & $9.25$ & $29$ & $17$ & $1\times 10^{-4}$ & $2.7\times 10^{11}$ & $1.6\times 10^{11}$ & $2\times 10^{22}$ & $1.6\times 10^{-22}$ & $9.6\times 10^{-23}$ \\
$$ & $\nu_{\alpha}\bar{\nu}_{\alpha}$ & $9.95$ & $29$ & $18$ & $4.4\times 10^{-4}$ & $6.7\times 10^{10}$ & $4.1\times 10^{10}$ & $2.2\times 10^{22}$ & $3.5\times 10^{-23}$ & $2.1\times 10^{-23}$ \\
&&&&&&&&&& \\[-0.1in]
\hline
&&&&&&&&&& \\[-0.1in]
$750$ & $b\bar{b}$ & $8.15$ & $27$ & $21$ & $4.4\times 10^{-7}$ & $6.1\times 10^{13}$ & $4.7\times 10^{13}$ & $1.8\times 10^{22}$ & $2.5\times 10^{-21}$ & $1.9\times 10^{-21}$ \\
$$ & $W^{+}W^{-}$ & $9.05$ & $28$ & $18$ & $5.4\times 10^{-6}$ & $5.2\times 10^{12}$ & $3.3\times 10^{12}$ & $2\times 10^{22}$ & $3.6\times 10^{-22}$ & $2.3\times 10^{-22}$ \\
$$ & $\tau\bar{\tau}$ & $9.05$ & $28$ & $18$ & $9.4\times 10^{-5}$ & $3\times 10^{11}$ & $1.9\times 10^{11}$ & $2\times 10^{22}$ & $1.8\times 10^{-22}$ & $1.1\times 10^{-22}$ \\
$$ & $\mu^{+}\mu^{-}$ & $9.05$ & $28$ & $18$ & $1.9\times 10^{-4}$ & $1.5\times 10^{11}$ & $9.3\times 10^{10}$ & $2\times 10^{22}$ & $1.8\times 10^{-22}$ & $1.1\times 10^{-22}$ \\
$$ & $\nu_{\alpha}\bar{\nu}_{\alpha}$ & $9.05$ & $28$ & $18$ & $8.5\times 10^{-4}$ & $3.3\times 10^{10}$ & $2.1\times 10^{10}$ & $2\times 10^{22}$ & $3.4\times 10^{-23}$ & $2.2\times 10^{-23}$ \\
&&&&&&&&&& \\[-0.1in]
\hline
&&&&&&&&&& \\[-0.1in]
$1000$ & $b\bar{b}$ & $8.05$ & $27$ & $20$ & $7.1\times 10^{-7}$ & $3.8\times 10^{13}$ & $2.8\times 10^{13}$ & $1.8\times 10^{22}$ & $2.5\times 10^{-21}$ & $1.8\times 10^{-21}$ \\
$$ & $W^{+}W^{-}$ & $8.05$ & $27$ & $20$ & $7.8\times 10^{-6}$ & $3.4\times 10^{12}$ & $2.6\times 10^{12}$ & $1.8\times 10^{22}$ & $4.4\times 10^{-22}$ & $3.3\times 10^{-22}$ \\
$$ & $\tau\bar{\tau}$ & $9.05$ & $28$ & $18$ & $1.5\times 10^{-4}$ & $1.9\times 10^{11}$ & $1.2\times 10^{11}$ & $2\times 10^{22}$ & $1.9\times 10^{-22}$ & $1.2\times 10^{-22}$ \\
$$ & $\mu^{+}\mu^{-}$ & $9.05$ & $28$ & $18$ & $3\times 10^{-4}$ & $9.3\times 10^{10}$ & $5.9\times 10^{10}$ & $2\times 10^{22}$ & $1.9\times 10^{-22}$ & $1.2\times 10^{-22}$ \\
$$ & $\nu_{\alpha}\bar{\nu}_{\alpha}$ & $8.15$ & $27$ & $21$ & $1.1\times 10^{-3}$ & $2.6\times 10^{10}$ & $2\times 10^{10}$ & $1.8\times 10^{22}$ & $4.5\times 10^{-23}$ & $3.5\times 10^{-23}$ \\
&&&&&&&&&& \\[-0.1in]
\hline
&&&&&&&&&& \\[-0.1in]
$1500$ & $b\bar{b}$ & $8.05$ & $27$ & $20$ & $1.4\times 10^{-6}$ & $2\times 10^{13}$ & $1.5\times 10^{13}$ & $1.8\times 10^{22}$ & $2.5\times 10^{-21}$ & $1.9\times 10^{-21}$ \\
$$ & $W^{+}W^{-}$ & $8.15$ & $27$ & $21$ & $1.4\times 10^{-5}$ & $1.9\times 10^{12}$ & $1.5\times 10^{12}$ & $1.8\times 10^{22}$ & $4.9\times 10^{-22}$ & $3.8\times 10^{-22}$ \\
$$ & $\tau\bar{\tau}$ & $8.05$ & $27$ & $20$ & $2.8\times 10^{-4}$ & $9.8\times 10^{10}$ & $7.3\times 10^{10}$ & $1.8\times 10^{22}$ & $2.4\times 10^{-22}$ & $1.8\times 10^{-22}$ \\
$$ & $\mu^{+}\mu^{-}$ & $8.05$ & $27$ & $20$ & $5.2\times 10^{-4}$ & $5.2\times 10^{10}$ & $3.9\times 10^{10}$ & $1.8\times 10^{22}$ & $2.4\times 10^{-22}$ & $1.8\times 10^{-22}$ \\
$$ & $\nu_{\alpha}\bar{\nu}_{\alpha}$ & $10.1$ & $30$ & $18$ & $1.9\times 10^{-3}$ & $1.5\times 10^{10}$ & $9.3\times 10^{09}$ & $2.2\times 10^{22}$ & $4.1\times 10^{-23}$ & $2.5\times 10^{-23}$ \\
&&&&&&&&&& \\[-0.1in]
\hline
&&&&&&&&&& \\[-0.1in]
$2000$ & $b\bar{b}$ & $8.05$ & $27$ & $20$ & $2.1\times 10^{-6}$ & $1.3\times 10^{13}$ & $9.3\times 10^{12}$ & $1.8\times 10^{22}$ & $2.6\times 10^{-21}$ & $2\times 10^{-21}$ \\
$$ & $W^{+}W^{-}$ & $8.75$ & $28$ & $18$ & $2\times 10^{-5}$ & $1.4\times 10^{12}$ & $9.2\times 10^{11}$ & $1.9\times 10^{22}$ & $5.6\times 10^{-22}$ & $3.6\times 10^{-22}$ \\
$$ & $\tau\bar{\tau}$ & $8.15$ & $27$ & $21$ & $4.1\times 10^{-4}$ & $6.6\times 10^{10}$ & $5.1\times 10^{10}$ & $1.8\times 10^{22}$ & $2.7\times 10^{-22}$ & $2.1\times 10^{-22}$ \\
$$ & $\mu^{+}\mu^{-}$ & $8.15$ & $27$ & $21$ & $7.4\times 10^{-4}$ & $3.6\times 10^{10}$ & $2.8\times 10^{10}$ & $1.8\times 10^{22}$ & $2.7\times 10^{-22}$ & $2.1\times 10^{-22}$ \\
$$ & $\nu_{\alpha}\bar{\nu}_{\alpha}$ & $8.75$ & $28$ & $18$ & $1.9\times 10^{-3}$ & $1.5\times 10^{10}$ & $9.5\times 10^{09}$ & $1.9\times 10^{22}$ & $7\times 10^{-23}$ & $4.5\times 10^{-23}$ \\

\hline
\hline
\end{tabular}
\caption{Extension of Table~\ref{tab:qresultsone} for $\rm M_{\rm WIMP}>360$ GeV.
\label{tab:qresultstwo}}
\end{footnotesize}
\end{center}
\end{table*}

\begin{table*}[!b]
\begin{center}
\begin{footnotesize}
\begin{tabular}{ccccccccccc}
\hline
\hline
&&&&&&&&&& \\[-0.1in]
$\rm M_{WIMP}$ & Channel & $\rm \Psi$ & $\rm \bar{\mu}^{90\%}$ & $\rm \mu^{90\%}$ & $\rm \bar{A}_{eff}(M_{\rm WIMP})$ & $\rm \overline{\Phi}_{\nu_{\mu}+\bar{\nu}_\mu}$ & $\rm \Phi_{\nu_{\mu}+\bar{\nu}_\mu}$ & $\rm J-Factor$ & $\rm \overline{<\sigma_{A}v>}$ & $\rm <\sigma_{A}v>$ \\
&&&&& $\rm \times T_{eff}$ &&&&& \\
$\rm (GeV)$ & & ($\rm ^{\circ}$) & & & ($\rm m^2.yr$) & ($\rm km^{-2}.yr^{-1}$) & ($\rm km^{-2}.yr^{-1}$) & ($\rm GeV^{2}.cm^{-5}$) & ($\rm cm^{3}.s^{-1}$) & ($\rm cm^{3}.s^{-1}$) \\
&&&&&&&&&& \\[-0.1in]

&&&&&&&&&& \\[-0.1in]
\hline
&&&&&&&&&& \\[-0.1in]
$25$ & $b\bar{b}$ & $7.05$ & $13$ & $14$ & $3.1\times 10^{-13}$ & $4.1\times 10^{19}$ & $4.4\times 10^{19}$ & $1.6\times 10^{22}$ & $6.8\times 10^{-18}$ & $7.3\times 10^{-18}$ \\
$$ & $\tau\bar{\tau}$ & $7.05$ & $13$ & $14$ & $5.7\times 10^{-10}$ & $2.2\times 10^{16}$ & $2.4\times 10^{16}$ & $1.6\times 10^{22}$ & $2.1\times 10^{-20}$ & $2.2\times 10^{-20}$ \\
$$ & $\mu^{+}\mu^{-}$ & $7.05$ & $13$ & $14$ & $1.6\times 10^{-9}$ & $7.9\times 10^{15}$ & $8.4\times 10^{15}$ & $1.6\times 10^{22}$ & $1.9\times 10^{-20}$ & $2\times 10^{-20}$ \\
$$ & $\nu_{\mu}\bar{\nu}_{\mu}$ & $7.05$ & $13$ & $14$ & $1.8\times 10^{-8}$ & $7\times 10^{14}$ & $7.5\times 10^{14}$ & $1.6\times 10^{22}$ & $2.2\times 10^{-21}$ & $2.4\times 10^{-21}$ \\
&&&&&&&&&& \\[-0.1in]
\hline
&&&&&&&&&& \\[-0.1in]
$50$ & $b\bar{b}$ & $14.6$ & $28$ & $36$ & $4\times 10^{-11}$ & $6.9\times 10^{17}$ & $9\times 10^{17}$ & $3\times 10^{22}$ & $1.9\times 10^{-19}$ & $2.5\times 10^{-19}$ \\
$$ & $\tau\bar{\tau}$ & $3.15$ & $6$ & $7.8$ & $8.4\times 10^{-9}$ & $7.2\times 10^{14}$ & $9.3\times 10^{14}$ & $7.5\times 10^{21}$ & $5.7\times 10^{-21}$ & $7.5\times 10^{-21}$ \\
$$ & $\mu^{+}\mu^{-}$ & $3.15$ & $6$ & $7.8$ & $2.2\times 10^{-8}$ & $2.7\times 10^{14}$ & $3.5\times 10^{14}$ & $7.5\times 10^{21}$ & $5.5\times 10^{-21}$ & $7.1\times 10^{-21}$ \\
$$ & $\nu_{\mu}\bar{\nu}_{\mu}$ & $2.45$ & $5$ & $9.5$ & $5.2\times 10^{-7}$ & $9.7\times 10^{12}$ & $1.8\times 10^{13}$ & $5.9\times 10^{21}$ & $3.3\times 10^{-22}$ & $6.3\times 10^{-22}$ \\
&&&&&&&&&& \\[-0.1in]
\hline
&&&&&&&&&& \\[-0.1in]
$90$ & $b\bar{b}$ & $2.55$ & $5.2$ & $9$ & $2.3\times 10^{-10}$ & $2.2\times 10^{16}$ & $3.9\times 10^{16}$ & $6.2\times 10^{21}$ & $8\times 10^{-20}$ & $1.4\times 10^{-19}$ \\
$$ & $W^{+}W^{-}$ & $2.55$ & $5.2$ & $9$ & $1.2\times 10^{-8}$ & $4.5\times 10^{14}$ & $7.8\times 10^{14}$ & $6.2\times 10^{21}$ & $1.9\times 10^{-21}$ & $3.3\times 10^{-21}$ \\
$$ & $\tau\bar{\tau}$ & $2.75$ & $5.3$ & $9.9$ & $3\times 10^{-7}$ & $1.8\times 10^{13}$ & $3.3\times 10^{13}$ & $6.7\times 10^{21}$ & $5.3\times 10^{-22}$ & $9.8\times 10^{-22}$ \\
$$ & $\mu^{+}\mu^{-}$ & $2.75$ & $5.3$ & $9.9$ & $8.2\times 10^{-7}$ & $6.5\times 10^{12}$ & $1.2\times 10^{13}$ & $6.7\times 10^{21}$ & $4.9\times 10^{-22}$ & $9\times 10^{-22}$ \\
$$ & $\nu_{\mu}\bar{\nu}_{\mu}$ & $1.85$ & $4.2$ & $3.5$ & $1.2\times 10^{-5}$ & $3.4\times 10^{11}$ & $2.9\times 10^{11}$ & $4.5\times 10^{21}$ & $5\times 10^{-23}$ & $4.2\times 10^{-23}$ \\
&&&&&&&&&& \\[-0.1in]
\hline
&&&&&&&&&& \\[-0.1in]
$100$ & $b\bar{b}$ & $2.55$ & $5.2$ & $9$ & $4\times 10^{-10}$ & $1.3\times 10^{16}$ & $2.3\times 10^{16}$ & $6.2\times 10^{21}$ & $5.6\times 10^{-20}$ & $9.7\times 10^{-20}$ \\
$$ & $W^{+}W^{-}$ & $2.75$ & $5.3$ & $9.9$ & $2.4\times 10^{-8}$ & $2.2\times 10^{14}$ & $4.2\times 10^{14}$ & $6.7\times 10^{21}$ & $1.1\times 10^{-21}$ & $2\times 10^{-21}$ \\
$$ & $\tau\bar{\tau}$ & $2.75$ & $5.3$ & $9.9$ & $5.2\times 10^{-7}$ & $1\times 10^{13}$ & $1.9\times 10^{13}$ & $6.7\times 10^{21}$ & $3.7\times 10^{-22}$ & $6.9\times 10^{-22}$ \\
$$ & $\mu^{+}\mu^{-}$ & $2.75$ & $5.3$ & $9.9$ & $1.4\times 10^{-6}$ & $3.8\times 10^{12}$ & $7\times 10^{12}$ & $6.7\times 10^{21}$ & $3.5\times 10^{-22}$ & $6.5\times 10^{-22}$ \\
$$ & $\nu_{\mu}\bar{\nu}_{\mu}$ & $1.85$ & $4.2$ & $3.5$ & $2.3\times 10^{-5}$ & $1.8\times 10^{11}$ & $\mathbf{1.5\times 10^{11}}$ & $4.5\times 10^{21}$ & $3.3\times 10^{-23}$ & $\mathbf{2.7\times 10^{-23}}$ \\
&&&&&&&&&& \\[-0.1in]
\hline
&&&&&&&&&& \\[-0.1in]
$150$ & $b\bar{b}$ & $2.75$ & $5.3$ & $9.9$ & $3.3\times 10^{-9}$ & $1.6\times 10^{15}$ & $3\times 10^{15}$ & $6.7\times 10^{21}$ & $1.3\times 10^{-20}$ & $2.3\times 10^{-20}$ \\
$$ & $W^{+}W^{-}$ & $3.05$ & $5.6$ & $9$ & $2.3\times 10^{-7}$ & $2.4\times 10^{13}$ & $\mathbf{3.9\times 10^{13}}$ & $7.3\times 10^{21}$ & $2.4\times 10^{-22}$ & $\mathbf{3.8\times 10^{-22}}$ \\
$$ & $\tau\bar{\tau}$ & $2.75$ & $5.3$ & $9.9$ & $3.1\times 10^{-6}$ & $1.7\times 10^{12}$ & $3.2\times 10^{12}$ & $6.7\times 10^{21}$ & $1.4\times 10^{-22}$ & $2.6\times 10^{-22}$ \\
$$ & $\mu^{+}\mu^{-}$ & $2.75$ & $5.3$ & $9.9$ & $8.6\times 10^{-6}$ & $6.2\times 10^{11}$ & $1.2\times 10^{12}$ & $6.7\times 10^{21}$ & $1.2\times 10^{-22}$ & $2.3\times 10^{-22}$ \\
$$ & $\nu_{\mu}\bar{\nu}_{\mu}$ & $1.55$ & $3.8$ & $2.7$ & $7.4\times 10^{-5}$ & $5.2\times 10^{10}$ & $\mathbf{3.6\times 10^{10}}$ & $3.8\times 10^{21}$ & $2.4\times 10^{-23}$ & $\mathbf{1.7\times 10^{-23}}$ \\
&&&&&&&&&& \\[-0.1in]
\hline
&&&&&&&&&& \\[-0.1in]
$180$ & $b\bar{b}$ & $2.75$ & $5.3$ & $9.9$ & $6.1\times 10^{-9}$ & $8.8\times 10^{14}$ & $1.6\times 10^{15}$ & $6.7\times 10^{21}$ & $9.2\times 10^{-21}$ & $1.7\times 10^{-20}$ \\
$$ & $W^{+}W^{-}$ & $3.05$ & $5.6$ & $9$ & $4.4\times 10^{-7}$ & $1.3\times 10^{13}$ & $\mathbf{2\times 10^{13}}$ & $7.3\times 10^{21}$ & $1.8\times 10^{-22}$ & $\mathbf{2.8\times 10^{-22}}$ \\
$$ & $\tau\bar{\tau}$ & $3.05$ & $5.6$ & $9$ & $5.7\times 10^{-6}$ & $1\times 10^{12}$ & $\mathbf{1.6\times 10^{12}}$ & $7.3\times 10^{21}$ & $1\times 10^{-22}$ & $\mathbf{1.7\times 10^{-22}}$ \\
$$ & $\mu^{+}\mu^{-}$ & $3.05$ & $5.6$ & $9$ & $1.5\times 10^{-5}$ & $3.7\times 10^{11}$ & $\mathbf{5.9\times 10^{11}}$ & $7.3\times 10^{21}$ & $9.5\times 10^{-23}$ & $\mathbf{1.5\times 10^{-22}}$ \\
$$ & $\nu_{\mu}\bar{\nu}_{\mu}$ & $2.85$ & $5.4$ & $9.7$ & $1.8\times 10^{-4}$ & $3\times 10^{10}$ & $\mathbf{5.5\times 10^{10}}$ & $6.9\times 10^{21}$ & $1\times 10^{-23}$ & $\mathbf{1.9\times 10^{-23}}$ \\
&&&&&&&&&& \\[-0.1in]
\hline
&&&&&&&&&& \\[-0.1in]
$200$ & $b\bar{b}$ & $2.75$ & $5.3$ & $9.9$ & $9.3\times 10^{-9}$ & $5.7\times 10^{14}$ & $1.1\times 10^{15}$ & $6.7\times 10^{21}$ & $7.2\times 10^{-21}$ & $1.3\times 10^{-20}$ \\
$$ & $W^{+}W^{-}$ & $3.05$ & $5.6$ & $9$ & $7.1\times 10^{-7}$ & $8\times 10^{12}$ & $\mathbf{1.3\times 10^{13}}$ & $7.3\times 10^{21}$ & $1.3\times 10^{-22}$ & $\mathbf{2.1\times 10^{-22}}$ \\
$$ & $\tau\bar{\tau}$ & $3.05$ & $5.6$ & $9$ & $8.7\times 10^{-6}$ & $6.5\times 10^{11}$ & $\mathbf{1\times 10^{12}}$ & $7.3\times 10^{21}$ & $8.3\times 10^{-23}$ & $\mathbf{1.3\times 10^{-22}}$ \\
$$ & $\mu^{+}\mu^{-}$ & $3.05$ & $5.6$ & $9$ & $2.3\times 10^{-5}$ & $2.4\times 10^{11}$ & $\mathbf{3.9\times 10^{11}}$ & $7.3\times 10^{21}$ & $7.6\times 10^{-23}$ & $\mathbf{1.2\times 10^{-22}}$ \\
$$ & $\nu_{\mu}\bar{\nu}_{\mu}$ & $3.05$ & $5.6$ & $9$ & $2.8\times 10^{-4}$ & $2\times 10^{10}$ & $\mathbf{3.3\times 10^{10}}$ & $7.3\times 10^{21}$ & $7.7\times 10^{-24}$ & $\mathbf{1.2\times 10^{-23}}$ \\
&&&&&&&&&& \\[-0.1in]
\hline
&&&&&&&&&& \\[-0.1in]
$260$ & $b\bar{b}$ & $2.75$ & $5.3$ & $9.9$ & $2.9\times 10^{-8}$ & $1.8\times 10^{14}$ & $3.4\times 10^{14}$ & $6.7\times 10^{21}$ & $3.5\times 10^{-21}$ & $6.5\times 10^{-21}$ \\
$$ & $W^{+}W^{-}$ & $2.25$ & $4.7$ & $9$ & $1.6\times 10^{-6}$ & $3\times 10^{12}$ & $\mathbf{5.7\times 10^{12}}$ & $5.4\times 10^{21}$ & $1.1\times 10^{-22}$ & $\mathbf{2.1\times 10^{-22}}$ \\
$$ & $\tau\bar{\tau}$ & $2.85$ & $5.4$ & $9.7$ & $2.2\times 10^{-5}$ & $2.4\times 10^{11}$ & $\mathbf{4.3\times 10^{11}}$ & $6.9\times 10^{21}$ & $5.5\times 10^{-23}$ & $\mathbf{9.9\times 10^{-23}}$ \\
$$ & $\mu^{+}\mu^{-}$ & $2.85$ & $5.4$ & $9.7$ & $5.9\times 10^{-5}$ & $9.1\times 10^{10}$ & $\mathbf{1.6\times 10^{11}}$ & $6.9\times 10^{21}$ & $4.9\times 10^{-23}$ & $\mathbf{8.9\times 10^{-23}}$ \\
$$ & $\nu_{\mu}\bar{\nu}_{\mu}$ & $1.85$ & $4.2$ & $3.5$ & $4\times 10^{-4}$ & $1\times 10^{10}$ & $\mathbf{8.7\times 10^{9}}$ & $4.5\times 10^{21}$ & $9.6\times 10^{-24}$ & $\mathbf{8.1\times 10^{-24}}$ \\
&&&&&&&&&& \\[-0.1in]
\hline
&&&&&&&&&& \\[-0.1in]
$360$ & $b\bar{b}$ & $2.85$ & $5.4$ & $9.7$ & $1.1\times 10^{-7}$ & $5.1\times 10^{13}$ & $\mathbf{9.2\times 10^{13}}$ & $6.9\times 10^{21}$ & $1.6\times 10^{-21}$ & $\mathbf{2.9\times 10^{-21}}$ \\
$$ & $W^{+}W^{-}$ & $1.85$ & $4.2$ & $3.5$ & $3.8\times 10^{-6}$ & $1.1\times 10^{12}$ & $\mathbf{9.2\times 10^{11}}$ & $4.5\times 10^{21}$ & $8.9\times 10^{-23}$ & $\mathbf{7.5\times 10^{-23}}$ \\
$$ & $\tau\bar{\tau}$ & $1.85$ & $4.2$ & $3.5$ & $5\times 10^{-5}$ & $8.3\times 10^{10}$ & $\mathbf{7\times 10^{10}}$ & $4.5\times 10^{21}$ & $5.4\times 10^{-23}$ & $\mathbf{4.5\times 10^{-23}}$ \\
$$ & $\mu^{+}\mu^{-}$ & $1.85$ & $4.2$ & $3.5$ & $1.3\times 10^{-4}$ & $3.2\times 10^{10}$ & $\mathbf{2.7\times 10^{10}}$ & $4.5\times 10^{21}$ & $4.9\times 10^{-23}$ & $\mathbf{4.1\times 10^{-23}}$ \\
$$ & $\nu_{\mu}\bar{\nu}_{\mu}$ & $1.85$ & $4.2$ & $3.5$ & $8.8\times 10^{-4}$ & $4.7\times 10^{9}$ & $\mathbf{4\times 10^{9}}$ & $4.5\times 10^{21}$ & $7.2\times 10^{-24}$ & $\mathbf{6\times 10^{-24}}$ \\

\hline
\hline
\end{tabular}
\caption{Equivalent of Table~\ref{tab:qresultsone} for $\rm \Lambda Fit$.
\label{tab:qresultsthree}}
\end{footnotesize}
\end{center}
\end{table*}

\begin{table*}[!b]
\begin{center}
\begin{footnotesize}
\begin{tabular}{ccccccccccc}
\hline
\hline
&&&&&&&&&& \\[-0.1in]
$\rm M_{WIMP}$ & Channel & $\rm \Psi$ & $\rm \bar{\mu}^{90\%}$ & $\rm \mu^{90\%}$ & $\rm \bar{A}_{eff}(M_{\rm WIMP})$ & $\rm \overline{\Phi}_{\nu_{\mu}+\bar{\nu}_\mu}$ & $\rm \Phi_{\nu_{\mu}+\bar{\nu}_\mu}$ & $\rm J-Factor$ & $\rm \overline{<\sigma_{A}v>}$ & $\rm <\sigma_{A}v>$ \\
&&&&& $\rm \times T_{eff}$ &&&&& \\
$\rm (GeV)$ & & ($\rm ^{\circ}$) & & & ($\rm m^2.yr$) & ($\rm km^{-2}.yr^{-1}$) & ($\rm km^{-2}.yr^{-1}$) & ($\rm GeV^{2}.cm^{-5}$) & ($\rm cm^{3}.s^{-1}$) & ($\rm cm^{3}.s^{-1}$) \\
&&&&&&&&&& \\[-0.1in]

&&&&&&&&&& \\[-0.1in]
\hline
&&&&&&&&&& \\[-0.1in]
$500$ & $b\bar{b}$ & $2.85$ & $5.4$ & $9.7$ & $2.3\times 10^{-7}$ & $2.4\times 10^{13}$ & $\mathbf{4.3\times 10^{13}}$ & $6.9\times 10^{21}$ & $1.3\times 10^{-21}$ & $\mathbf{2.4\times 10^{-21}}$ \\
$$ & $W^{+}W^{-}$ & $1.85$ & $4.2$ & $3.5$ & $7.1\times 10^{-6}$ & $5.9\times 10^{11}$ & $\mathbf{5\times 10^{11}}$ & $4.5\times 10^{21}$ & $8.7\times 10^{-23}$ & $\mathbf{7.4\times 10^{-23}}$ \\
$$ & $\tau\bar{\tau}$ & $1.85$ & $4.2$ & $3.5$ & $9.7\times 10^{-5}$ & $4.3\times 10^{10}$ & $\mathbf{3.6\times 10^{10}}$ & $4.5\times 10^{21}$ & $5.2\times 10^{-23}$ & $\mathbf{4.4\times 10^{-23}}$ \\
$$ & $\mu^{+}\mu^{-}$ & $1.85$ & $4.2$ & $3.5$ & $2.4\times 10^{-4}$ & $1.8\times 10^{10}$ & $\mathbf{1.5\times 10^{10}}$ & $4.5\times 10^{21}$ & $4.7\times 10^{-23}$ & $\mathbf{4\times 10^{-23}}$ \\
$$ & $\nu_{\mu}\bar{\nu}_{\mu}$ & $1.05$ & $2.9$ & $2$ & $1\times 10^{-3}$ & $2.8\times 10^{9}$ & $\mathbf{2\times 10^{9}}$ & $2.5\times 10^{21}$ & $1.2\times 10^{-23}$ & $\mathbf{8.6\times 10^{-24}}$ \\
&&&&&&&&&& \\[-0.1in]
\hline
&&&&&&&&&& \\[-0.1in]
$750$ & $b\bar{b}$ & $1.85$ & $4.2$ & $3.5$ & $6.3\times 10^{-7}$ & $6.6\times 10^{12}$ & $\mathbf{5.6\times 10^{12}}$ & $4.5\times 10^{21}$ & $1.1\times 10^{-21}$ & $\mathbf{9.1\times 10^{-22}}$ \\
$$ & $W^{+}W^{-}$ & $1.05$ & $2.9$ & $2$ & $1.3\times 10^{-5}$ & $2.3\times 10^{11}$ & $\mathbf{1.6\times 10^{11}}$ & $2.5\times 10^{21}$ & $1.2\times 10^{-22}$ & $\mathbf{8.6\times 10^{-23}}$ \\
$$ & $\tau\bar{\tau}$ & $1.85$ & $4.2$ & $3.5$ & $3\times 10^{-4}$ & $1.4\times 10^{10}$ & $\mathbf{1.2\times 10^{10}}$ & $4.5\times 10^{21}$ & $3.7\times 10^{-23}$ & $\mathbf{3.1\times 10^{-23}}$ \\
$$ & $\mu^{+}\mu^{-}$ & $1.85$ & $4.2$ & $3.5$ & $6.8\times 10^{-4}$ & $6.1\times 10^{9}$ & $\mathbf{5.1\times 10^{9}}$ & $4.5\times 10^{21}$ & $3.4\times 10^{-23}$ & $\mathbf{2.8\times 10^{-23}}$ \\
$$ & $\nu_{\mu}\bar{\nu}_{\mu}$ & $1.05$ & $2.9$ & $2$ & $2\times 10^{-3}$ & $1.4\times 10^{9}$ & $\mathbf{9.9\times 10^{8}}$ & $2.5\times 10^{21}$ & $1.2\times 10^{-23}$ & $\mathbf{8.1\times 10^{-24}}$ \\
&&&&&&&&&& \\[-0.1in]
\hline
&&&&&&&&&& \\[-0.1in]
$1000$ & $b\bar{b}$ & $1.85$ & $4.2$ & $3.5$ & $1.5\times 10^{-6}$ & $2.8\times 10^{12}$ & $\mathbf{2.4\times 10^{12}}$ & $4.5\times 10^{21}$ & $7.5\times 10^{-22}$ & $\mathbf{6.3\times 10^{-22}}$ \\
$$ & $W^{+}W^{-}$ & $1.05$ & $2.9$ & $2$ & $2.7\times 10^{-5}$ & $1.1\times 10^{11}$ & $\mathbf{7.4\times 10^{10}}$ & $2.5\times 10^{21}$ & $9.7\times 10^{-23}$ & $\mathbf{6.7\times 10^{-23}}$ \\
$$ & $\tau\bar{\tau}$ & $1.05$ & $2.9$ & $2$ & $4.4\times 10^{-4}$ & $6.5\times 10^{9}$ & $\mathbf{4.5\times 10^{9}}$ & $2.5\times 10^{21}$ & $5.3\times 10^{-23}$ & $\mathbf{3.7\times 10^{-23}}$ \\
$$ & $\mu^{+}\mu^{-}$ & $1.05$ & $2.9$ & $2$ & $9.7\times 10^{-4}$ & $3\times 10^{9}$ & $\mathbf{2.1\times 10^{9}}$ & $2.5\times 10^{21}$ & $4.8\times 10^{-23}$ & $\mathbf{3.4\times 10^{-23}}$ \\
$$ & $\nu_{\mu}\bar{\nu}_{\mu}$ & $1.05$ & $2.9$ & $2$ & $4.4\times 10^{-3}$ & $6.6\times 10^{8}$ & $\mathbf{4.6\times 10^{8}}$ & $2.5\times 10^{21}$ & $8.3\times 10^{-24}$ & $\mathbf{5.8\times 10^{-24}}$ \\
&&&&&&&&&& \\[-0.1in]
\hline
&&&&&&&&&& \\[-0.1in]
$1500$ & $b\bar{b}$ & $1.85$ & $4.2$ & $3.5$ & $3.4\times 10^{-6}$ & $1.2\times 10^{12}$ & $\mathbf{1\times 10^{12}}$ & $4.5\times 10^{21}$ & $6.4\times 10^{-22}$ & $\mathbf{5.4\times 10^{-22}}$ \\
$$ & $W^{+}W^{-}$ & $1.05$ & $2.9$ & $2$ & $5.2\times 10^{-5}$ & $5.6\times 10^{10}$ & $\mathbf{3.9\times 10^{10}}$ & $2.5\times 10^{21}$ & $1\times 10^{-22}$ & $7\mathbf{.1\times 10^{-23}}$ \\
$$ & $\tau\bar{\tau}$ & $1.05$ & $2.9$ & $2$ & $9.6\times 10^{-4}$ & $3\times 10^{9}$ & $\mathbf{2.1\times 10^{9}}$ & $2.5\times 10^{21}$ & $5.2\times 10^{-23}$ & $\mathbf{3.6\times 10^{-23}}$ \\
$$ & $\mu^{+}\mu^{-}$ & $1.05$ & $2.9$ & $2$ & $2\times 10^{-3}$ & $1.5\times 10^{9}$ & $\mathbf{1\times 10^{9}}$ & $2.5\times 10^{21}$ & $4.8\times 10^{-23}$ & $\mathbf{3.3\times 10^{-23}}$ \\
$$ & $\nu_{\mu}\bar{\nu}_{\mu}$ & $1.05$ & $2.9$ & $2$ & $6.5\times 10^{-3}$ & $4.4\times 10^{8}$ & $\mathbf{3.1\times 10^{8}}$ & $2.5\times 10^{21}$ & $1\times 10^{-23}$ & $\mathbf{7.2\times 10^{-24}}$ \\
&&&&&&&&&& \\[-0.1in]
\hline
&&&&&&&&&& \\[-0.1in]
$2000$ & $b\bar{b}$ & $1.85$ & $4.2$ & $3.5$ & $7\times 10^{-6}$ & $6\times 10^{11}$ & $\mathbf{5\times 10^{11}}$ & $4.5\times 10^{21}$ & $5\times 10^{-22}$ & $\mathbf{4.2\times 10^{-22}}$ \\
$$ & $W^{+}W^{-}$ & $0.95$ & $2.7$ & $2.2$ & $9.2\times 10^{-5}$ & $2.9\times 10^{10}$ & $\mathbf{2.4\times 10^{10}}$ & $2.3\times 10^{21}$ & $9.8\times 10^{-23}$ & $\mathbf{8\times 10^{-23}}$ \\
$$ & $\tau\bar{\tau}$ & $1.05$ & $2.9$ & $2$ & $1.9\times 10^{-3}$ & $1.5\times 10^{9}$ & $\mathbf{1\times 10^{9}}$ & $2.5\times 10^{21}$ & $4.4\times 10^{-23}$ & $\mathbf{3\times 10^{-23}}$ \\
$$ & $\mu^{+}\mu^{-}$ & $1.05$ & $2.9$ & $2$ & $3.8\times 10^{-3}$ & $7.5\times 10^{8}$ & $\mathbf{5.2\times 10^{8}}$ & $2.5\times 10^{21}$ & $4.1\times 10^{-23}$ & $\mathbf{2.8\times 10^{-23}}$ \\
$$ & $\nu_{\mu}\bar{\nu}_{\mu}$ & $0.95$ & $2.7$ & $2.2$ & $1.1\times 10^{-3}$ & $2.4\times 10^{8}$ & $\mathbf{1.9\times 10^{8}}$ & $2.3\times 10^{21}$ & $9.5\times 10^{-24}$ & $\mathbf{7.7\times 10^{-24}}$ \\
&&&&&&&&&& \\[-0.1in]
\hline
&&&&&&&&&& \\[-0.1in]
$3000$ & $b\bar{b}$ & $1.05$ & $2.9$ & $2$ & $1.5\times 10^{-5}$ & $2\times 10^{11}$ & $\mathbf{1.4\times 10^{11}}$ & $2.5\times 10^{21}$ & $5.8\times 10^{-22}$ & $\mathbf{4\times 10^{-22}}$ \\
$$ & $W^{+}W^{-}$ & $0.95$ & $2.7$ & $2.2$ & $2.4\times 10^{-4}$ & $1.1\times 10^{10}$ & $\mathbf{9.1\times 10^{9}}$ & $2.3\times 10^{21}$ & $7.5\times 10^{-23}$ & $\mathbf{6.1\times 10^{-23}}$ \\
$$ & $\tau\bar{\tau}$ & $0.95$ & $2.7$ & $2.2$ & $5.2\times 10^{-3}$ & $5.2\times 10^{8}$ & $\mathbf{4.2\times 10^{8}}$ & $2.3\times 10^{21}$ & $3.6\times 10^{-23}$ & $\mathbf{2.9\times 10^{-23}}$ \\
$$ & $\mu^{+}\mu^{-}$ & $0.95$ & $2.7$ & $2.2$ & $9.6\times 10^{-3}$ & $2.8\times 10^{8}$ & $\mathbf{2.3\times 10^{8}}$ & $2.3\times 10^{21}$ & $3.4\times 10^{-23}$ & $\mathbf{2.8\times 10^{-23}}$ \\
$$ & $\nu_{\mu}\bar{\nu}_{\mu}$ & $0.95$ & $2.7$ & $2.2$ & $2.8\times 10^{-3}$ & $9.5\times 10^{7}$ & $\mathbf{7.7\times 10^{7}}$ & $2.3\times 10^{21}$ & $7.3\times 10^{-24}$ & $\mathbf{5.9\times 10^{-24}}$ \\
&&&&&&&&&& \\[-0.1in]
\hline
&&&&&&&&&& \\[-0.1in]
$5000$ & $b\bar{b}$ & $1.05$ & $2.9$ & $2$ & $5.7\times 10^{-5}$ & $5\times 10^{10}$ & $\mathbf{3.5\times 10^{10}}$ & $2.5\times 10^{21}$ & $3.5\times 10^{-22}$ & $\mathbf{2.4\times 10^{-22}}$ \\
$$ & $W^{+}W^{-}$ & $0.95$ & $2.7$ & $2.2$ & $7.4\times 10^{-4}$ & $3.7\times 10^{9}$ & $\mathbf{3\times 10^{9}}$ & $2.3\times 10^{21}$ & $5.8\times 10^{-23}$ & $\mathbf{4.7\times 10^{-23}}$ \\
$$ & $\tau\bar{\tau}$ & $0.95$ & $2.7$ & $2.2$ & $1.9\times 10^{-2}$ & $1.4\times 10^{8}$ & $\mathbf{1.1\times 10^{8}}$ & $2.3\times 10^{21}$ & $2.5\times 10^{-23}$ & $\mathbf{2\times 10^{-23}}$ \\
$$ & $\mu^{+}\mu^{-}$ & $0.95$ & $2.7$ & $2.2$ & $3.4\times 10^{-2}$ & $8.1\times 10^{7}$ & $\mathbf{6.5\times 10^{7}}$ & $2.3\times 10^{21}$ & $2.4\times 10^{-23}$ & $\mathbf{1.9\times 10^{-23}}$ \\
$$ & $\nu_{\mu}\bar{\nu}_{\mu}$ & $0.95$ & $2.7$ & $2.2$ & $7.1\times 10^{-2}$ & $3.8\times 10^{7}$ & $\mathbf{3.1\times 10^{7}}$ & $2.3\times 10^{21}$ & $6.6\times 10^{-24}$ & $\mathbf{5.3\times 10^{-24}}$ \\
&&&&&&&&&& \\[-0.1in]
\hline
&&&&&&&&&& \\[-0.1in]
$10000$ & $b\bar{b}$ & $0.95$ & $2.7$ & $2.2$ & $3.3\times 10^{-4}$ & $8.2\times 10^{9}$ & $\mathbf{6.6\times 10^{9}}$ & $2.3\times 10^{21}$ & $2\times 10^{-22}$ & $\mathbf{1.6\times 10^{-22}}$ \\
$$ & $W^{+}W^{-}$ & $0.95$ & $2.7$ & $2.2$ & $3.2\times 10^{-3}$ & $8.4\times 10^{8}$ & $\mathbf{6.8\times 10^{8}}$ & $2.3\times 10^{21}$ & $4.3\times 10^{-23}$ & $\mathbf{3.5\times 10^{-23}}$ \\
$$ & $\tau\bar{\tau}$ & $0.95$ & $2.7$ & $2.2$ & $1\times 10^{-1}$ & $2.6\times 10^{7}$ & $\mathbf{2.1\times 10^{7}}$ & $2.3\times 10^{21}$ & $1.7\times 10^{-23}$ & $\mathbf{1.4\times 10^{-23}}$ \\
$$ & $\mu^{+}\mu^{-}$ & $0.95$ & $2.7$ & $2.2$ & $1.6\times 10^{-1}$ & $1.7\times 10^{7}$ & $\mathbf{1.4\times 10^{7}}$ & $2.3\times 10^{21}$ & $1.7\times 10^{-23}$ & $\mathbf{1.4\times 10^{-23}}$ \\
$$ & $\nu_{\mu}\bar{\nu}_{\mu}$ & $0.95$ & $2.7$ & $2.2$ & $2.6\times 10^{-1}$ & $1\times 10^{7}$ & $\mathbf{8.5\times 10^{6}}$ & $2.3\times 10^{21}$ & $5.6\times 10^{-24}$ & $\mathbf{4.5\times 10^{-24}}$ \\

\hline
\hline
\end{tabular}
\caption{Extension of Table~\ref{tab:qresultsthree} for $\rm M_{\rm WIMP}>360$ GeV.
\label{tab:qresultsfour}}
\end{footnotesize}
\end{center}
\end{table*}

\end{document}